\documentclass[10pt,letterpaper]{article}
\usepackage[top=0.85in,left=1.1in,footskip=0.75in]{geometry}

\usepackage{floatrow}           
\usepackage{graphicx}
\usepackage{dcolumn}
\usepackage{array}
\usepackage{url}
\newcolumntype{P}[1]{>{\centering\arraybackslash}p{#1}} 
\newcolumntype{R}[1]{>{\raggedright\arraybackslash}p{#1}} 

\usepackage{amsmath,amssymb}
\usepackage{changepage}
\usepackage[utf8x]{inputenc}
\usepackage{bbm}
\usepackage[noadjust]{cite}
\usepackage[right]{lineno} 
\usepackage{microtype}
\DisableLigatures[f]{encoding = *, family = * }

\newlength\savedwidth

\usepackage[aboveskip=1pt,labelfont=bf,labelsep=period,justification=raggedright,singlelinecheck=off,figurename=Fig]{caption}

\makeatletter
\renewcommand{\@biblabel}[1]{\quad#1.}
\makeatother

\date{}

\usepackage{lastpage,fancyhdr}
\usepackage[]{graphicx} 
\usepackage[outdir=./]{epstopdf}
\usepackage{bm}
\usepackage{color}
\usepackage{bbold}
\usepackage{soul}
\usepackage{epsfig,epic}
\newcommand{\newc}{\newcommand}
\newc{\beq}{\begin{equation}}
\newc{\eeq}{\end{equation}}
\newc{\kt}{\rangle}
\newc{\br}{\langle}
\newc{\beqa}{\begin{eqnarray}}
\newc{\eeqa}{\end{eqnarray}}
\newc{\longra}{\longrightarrow}
\renewcommand{\eqref}[1]{Eq.~(\ref{eq:#1})}

\newcommand{\figref}[1]{Fig.~\ref{fig:#1}}
\newcommand{\siref}[1]{SI Text~\ref{sec:#1}}
\newcommand{\tabref}[1]{Tab.~\ref{tab:#1}}
\newcommand{\mutb}{u}
\newcommand{\mutm}{v}
\newcommand{\rr}{\lambda}
\newcommand{\pprog}{\gamma}
\newcommand{\Pb}{P_a}
\newcommand{\Pm}{P_c}
\newcommand{\C}{K}
\newcommand{\R}{R}
\renewcommand{\t}{\tilde{t}}
\renewcommand{\S}{A}

\renewcommand{\phi}{\varphi}

\begin{document}
\vspace*{0.2in}

\begin{flushleft}
{\Large
\textbf{Modeling age-specific incidence of colon cancer via niche
  competition}\\[1mm]
}

Steffen Lange\textsuperscript{1,2*},
Richard Mogwitz\textsuperscript{2},
Denis H\"unniger\textsuperscript{1,2},
Anja Vo\ss{}-B\"ohme\textsuperscript{1,2}
\\
\bigskip
\textbf{1} DataMedAssist, HTW Dresden, 01069 Dresden, Germany
\\
\textbf{2} Faculty of Informatics/Mathematics, HTW Dresden - University
  of Applied Sciences, 01069 Dresden
  \\
  \bigskip

* steffen.lange@tu-dresden.de\\
\end{flushleft}

\section*{Abstract}

Cancer development is a multistep process often starting with a single
cell in which a number of epigenetic and genetic alterations have
accumulated thus transforming it into a tumor cell. The progeny of
such a single benign tumor cell expands in the tissue and can at some
point progress to malignant tumor cells until a detectable tumor is
formed. The dynamics from the early phase of a single cell to a
detectable tumor with billions of tumor cells are complex and still
not fully resolved, not even for the well-known prototype of
multistage carcinogenesis, the adenoma-adenocarcinoma sequence of
colorectal cancer. Mathematical models of such carcinogenesis are
frequently tested and calibrated based on reported age-specific
incidence rates of cancer, but they usually require calibration of
four or more parameters due to the wide range of processes these
models aim to reflect. We present a cell-based model, which focuses on
the competition between wild-type and tumor cells in colonic crypts,
with which we are able reproduce epidemiological incidence rates of
colon cancer. Additionally, the fraction of cancerous tumors with
precancerous lesions predicted by the model agree with clinical
estimates. The correspondence between model and reported data suggests
that the fate of tumor development is majorly determined by the early
phase of tumor growth and progression long before a tumor becomes
detectable. Due to the focus on the early phase of tumor development,
the model has only a single fit parameter, the time scale set by an
effective replacement rate of stem cells in the crypt. We find this
effective rate to be considerable smaller than the actual replacement
rate, which implies that the time scale is limited by the processes
succeeding clonal conversion of crypts.

\section*{Introduction}

Cancer development is a multistep
process~\cite{VogKin1993,LitVinLi2008} often originating from a single
mutated cell~\cite{FeaHamVog1987}. Potential tumor progenitor cells in
the tissue accumulate sequentially epigenetic and genetic alterations,
which transform them into tumor cells~\cite{ReyMorClaWei2001}.
Initially, these tumor cells can be
benign~\cite{WilWerBarGraSot2016,WuWanLinLu2016}, meaning that they do
not possess a proliferative fitness advantage, and consequently
compete with the original wild-type cells within normal tissue
homeostasis~\cite{AmoBac2014}. When the first tumor cell acquires a
sufficient number of alterations and progresses to a malignant type,
i.e., gains a considerable proliferative advantage to the original
wild-type cells, a cancer develops~\cite{LodBerZipMatBalDar2000}: The
progeny of the malignant tumor cell spreads via clonal expansion until
a sufficiently large cell population is reached to be clinically
detectable~\cite{KuaNagEik2016}. A well-known prototype of such a
multistage carcinogenesis is the adenoma-adenocarcinoma sequence of
colorectal cancer~\cite{FoxWan2007,ZeuTodStaDeM2014}, whose
intra-tumor heterogeneity suggests that this cancer particularly
arises as a single expansion event~\cite{Sotetal2015}.

Besides this general framework of carcinogenesis, the exact processes
by which a tumor develops in the early phase are not known as a tumor
is usually only detected after it consists of billions of cells.
Mathematical models have been extensively used to elucidate
fundamental mechanisms of cancer development and progression on the
basis of biological data. One frequently employed interface to link
these dynamical models to real-world observations are age-specific
cancer incidences from cancer registries. Starting with the multistage
Armitage-Doll model~\cite{ArmDol1954,ArmDol2004,MooMezTur2009},
suggesting that cancer generation is governed by a sequence of
rate-limiting events, and multistage clonal expansion model
(MSCE)~\cite{MezJeoMooLue2008,MooMezTur2009,MezJeoRenLue2010,LueCurJeoHaz2013,MezCha2015,BroEisMez2016,BroMezEis2017},
which are based on the initiation-promotion-malignant conversion
paradigm in carcinogenesis, multitype branching process
models~\cite{CalTavShi2004,KimCalTavShi2004,LanKuiMisBee2020}, frailty
models~\cite{GroBraHolHauKunTreAalMog2011} as well as other stochastic
or regression
models~\cite{MdzGleKinShe2009,MdzShe2010,Bro2011,SotBro2012,RhyOhKimKimRhyHon2021}
have been applied to age-specific incidences of various types of
cancer. This includes
colon~\cite{CalTavShi2004,KimCalTavShi2004,LitVinLi2008,MezJeoRenLue2010,SotBro2012,LanKuiMisBee2020}
and colorectal
cancer~\cite{RhyOhKimKimRhyHon2021,LueCurJeoHaz2013,MezJeoMooLue2008,CalTavShi2004},
pancreatic
cancer~\cite{MdzGleKinShe2009,BroMezEis2017,LueCurJeoHaz2013,MezJeoMooLue2008},
gastric cancer~\cite{RhyOhKimKimRhyHon2021,LueCurJeoHaz2013},
esophageal adenocarcinomas (EAC)~\cite{LueCurJeoHaz2013}, oral
squamous cell carcinomas (OSCCs)~\cite{BroEisMez2016}, prostate
cancer~\cite{SotBro2012}, gonadal germ cell cancer (most common form
of testicular cancer)~\cite{Bro2011}, lung cancer~\cite{MdzShe2010},
kidney cancer~\cite{MdzGleKinShe2009}, thyroid
cancer~\cite{MezCha2015} as well as Hodgkin lymphoma
(HL)~\cite{GroBraHolHauKunTreAalMog2011}. Parameters of these models
are determined by fitting the models hazard function to the
age-specific incidences. A match of the fit and the data is considered
to support the validity of the corresponding model. Furthermore,
important parameter combinations, which quantify kinetics of malignant
progression or clonal expansion before clinical detection, such as
growth rates of adenoma, malignant transformation rates, extinction
probabilities, as well as sojourn or dwell times, can be estimated via
maximum likelihood methods from the fits. While the listed models fit
the epidemiological data exquisitely well, they have at least
two~\cite{RhyOhKimKimRhyHon2021} but usually
four~\cite{MezJeoRenLue2010,MdzShe2010,Bro2011,SotBro2012,LueCurJeoHaz2013,BroEisMez2016,RhyOhKimKimRhyHon2021}
or
more~\cite{LitVinLi2008,MezJeoMooLue2008,GroBraHolHauKunTreAalMog2011,LanKuiMisBee2020}
parameters. Thus, there may be issues of identifiability of the
parameters from the data for many of these
models~\cite{BroMezEis2017}, although the issue is well understood
mathematically. The large amount of fit parameters results from the
wide range of mechanisms these generic models aim to reflect, often
incorporating both the early phase of tumor initiation and the
early-to-late phase of clonal expansion. In contrast, the spatial
structure of the tissue from which the tumor originates is often
neglected, although tissue architecture is known to be crucial for
tumor evolution~\cite{NobBurLeLemVioKatBee2022}. A recent counter
example is a model of colorectal cancer
initiation~\cite{PatCleBoz2020}, which did not infer the age-specific
incidences rates, but recovered the lifetime risk of colorectal
adenomacarcinoma, while setting all parameters by experimentally
measured rates.
\begin{figure}[t!]
  \centering
\hspace*{-1cm}\includegraphics{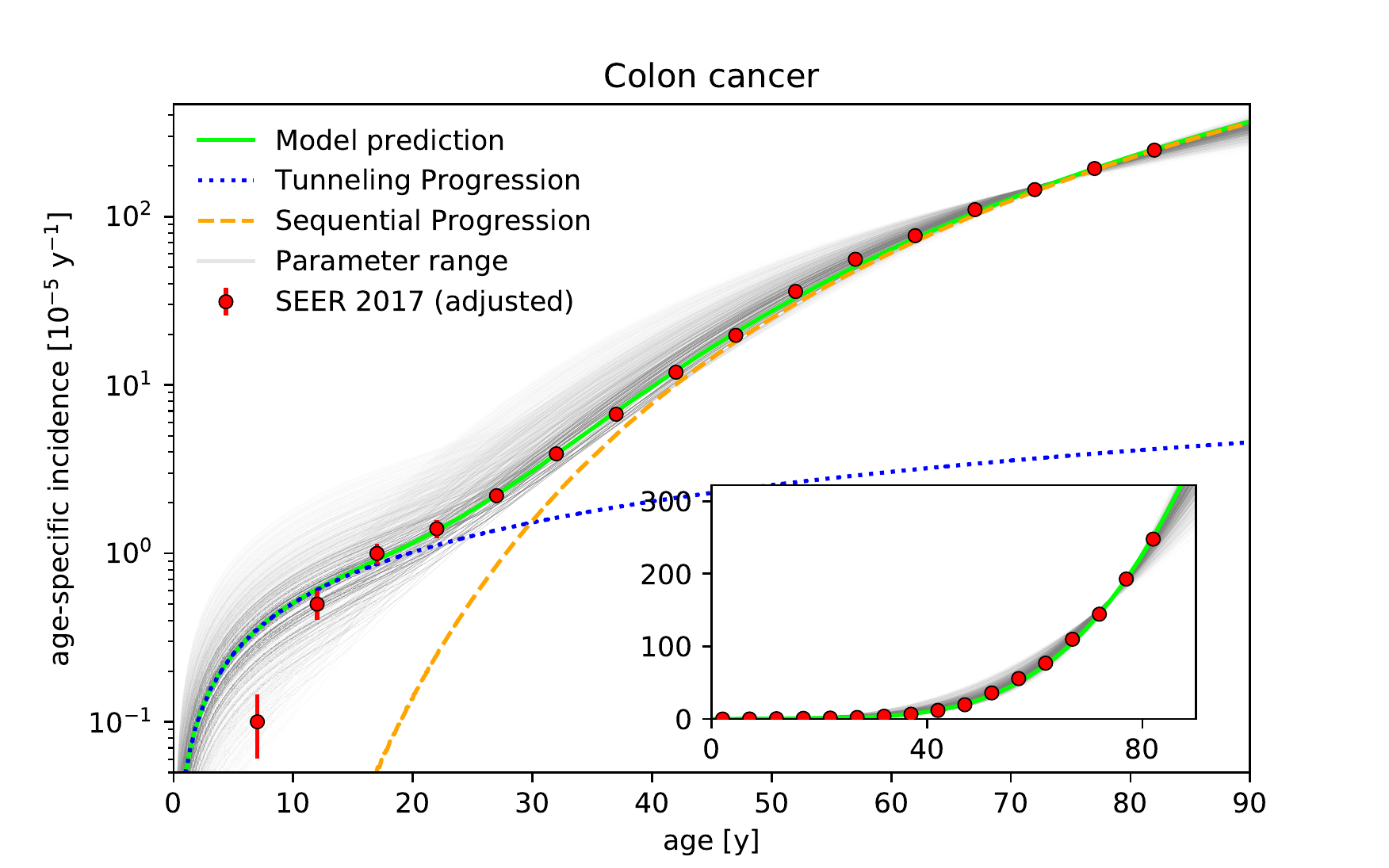}
\caption{\label{fig:incidence} \textbf{Niche competition in colonic
    crypts models age-specific incidence rates of colon cancer.}
  Age-specific incidence rates of colon cancer predicted by the model
  (green and gray lines) agree with epidemiological data (red dots,
  SEER database~\cite{SEER2017} adjusted for effects of colorectal
  screening at $>55$ years, see \siref{adjust}, error bars correspond
  to $95\%$ confidence interval assuming Poisson distribution and are
  often smaller than symbol size) over several orders of magnitude
  (see inset for linear y-scale). For the model prediction the range
  of the parameters niche size $N$, number of crypts $\C$, mutation
  probabilities $\mutb, \mutm$, and probability $\pprog$ of adenoma
  progressing to adenocarcinoma are taken from the literature, see
  \tabref{parameters}, while the time scale set by the effective
  replacement rate $\rr$ is calibrated based on the epidemiological
  data. Sensitivity on parameters is illustrated by variation of the
  parameters within the reported ranges (gray lines, opacity scaled
  inversely with corresponding goodness-of-fit $\sim 1/\chi^2$ for
  clarity). Exemplary parameter set with $N = 8$, $\C = 2 \cdot 10^7$,
  $\mutb=7.13\cdot10^{-6}$, $\mutm = 1.75\cdot10^{-6}$, and
  $\pprog = 9.4\%$ is highlighted in green. Decomposition of incidence
  rates predicted by the model into incidences with and without
  precancerous lesions (sequential and tunneling progression, orange
  dashed and dotted blue line respectively for the exemplary parameter
  set) confirms different origins of cancer at young and older ages.
  The effective replacement rate $\rr$ is found to be considerably
  smaller than the actual replacement rates
  ($\rr = 0.016 \text{ y}^{-1}$ per stem cell for the green curve and
  $\rr = 0.024 \pm 0.01\text{ y}^{-1}$ for the gray curves).}
\end{figure}

Our goal is to demonstrate that the competition between wild-type and
tumor cells in niches in the early phase of tumor initiation may be a
major mechanism for the fate of tumor development. For this, we
develop a cell-based stochastic model which represents the pretumor
competition between wild-type and tumor stem cells in colonic crypts
and thus both focuses on the early-phase of tumor initiation as well
as explicitly incorporates the spatial structure of the tissue. The
model reproduces the age-specific incidences of colon cancer from the
Surveillance Epidemiology and End Results (SEER) database
quantitatively, see \figref{incidence}. The model predictions agree
with the epidemiological data not only in the regime of older ages but
also down to young ages, where incidences are several orders of
magnitude smaller. This correspondence supports the recently proposed
notion that the fate of tumor development may be determined in the
early phase of tumor development long before a tumor becomes
detectable~\cite{BudDeuKliVos2019}. Since the model explicitly
distinguishes between (i) adenocarcinoma and (ii) adenoma, which
progressed to adenocarcinoma, we additionally predict the fraction of
incidences corresponding to either type. These fractions are in
agreement with the common clinical estimates that more than $95\%$ of
colon adenocarcinomas arise from colonic polyps and that the fraction
of benign tumor is over $99\%$~\cite{DriRid1994}. In particular, we
quantify how the fraction of incidences resulting from progressed
adenoma increases with age, which supports the relevance of colorectal
cancer screening at older age. We emphasize that all parameters of the
model correspond directly to known physiological parameters and are
consequently set by previously reported values. Only the effective
stem cell replacement rate within a crypt is used as a single fit
parameter to set the time scale. We find this effective replacement
rate to be considerably smaller than the actual replacement rate,
which implies that the time scale is limited by the processes
succeeding clonal conversion of a crypt.

\section*{Materials and Methods}

\subsection*{Cancer screening data}

The Surveillance Epidemiology and End Results (SEER) research database
comprises cancer incidences and at-risk population data in the
US~\cite{SEER2017}. We consider age-specific incidence rates of colon
cancer as reported in the most recent SEER dataset 2013-2017. For the
age-specific incidence rates in the SEER report the number of
diagnosed cancers, as a function of age, is compiled and then divided
by the corresponding total population at risk. The incidence rates are
reported in eighteen 5-year age groups and we assign rates to the
midpoint of these groups. The last age group $85+$ is excluded due to
rapidly declining person-years after age $85$. We estimate the
confidence interval of the rates by assuming a Poisson distribution of
the incidences and compute the range containing $95\%$ of the
incidences based on the US Standard population 2000~\cite{SEERpop} in
each age group.

However, the main source of uncertainty is not the statistical error
but the secular trends of incidence rates observed for colorectal
cancer: While the overall incidence of colorectal cancer has been
decreasing since 1998 due to a decrease in the age groups $>50$ years,
the incidence for men and women younger than $50$ has been rising, in
particular for adolescents and young
adults~\cite{SieDeSJem2014,Baietal2015}. The incidence rates exhibit
these age-group-dependent trends both over calendar year and birth
cohort, see also \figref{trend}. The decrease is attributed to the
widespread uptake of colonoscopy screening (and consequent resections)
for people above
$50$~\cite{KahImpJulRex2009,SieDeSJem2014,Levetal2018,CarZhuGuoHeiHofBre2021}.
To account for the effect of colorectal cancer screening, we consider
several reasonable adjustments of the incidence rates from the SEER
2017 dataset. As representative case, we display the results for colon
and rectum in \figref{incidence} and \figref{rectal-cancer},
respectively, for which the rates of all age groups above $55$ years
have been increased by a factor $4/3$. This adjustment is based on
estimates that the annual incidence of colorectal cancer at ages $>50$
between 2000 and 2015 is reduced on average by $\sim 25\%$ due to
screening~\cite{BreAltStoHof2015,Levetal2018}. Results for other
adjustments (including the case of the unaltered rates) are reported
in \siref{adjust}. For the colon, we find for each adjustment a
similarly good correspondence between the epidemiological incidence
rates and our model prediction. Mostly, the fitted effective
replacement rate for a particular parameter set is modulated by the
choice of the adjustment. Furthermore, we find agreement of our model
predictions with incidence rates of gastric cancer, whose secular
trends are far less pronounced.

At ages younger than $50$ colorectal screening occurs rarely. While
the upward trend of the incidences at these younger ages is less
understood, two opposing effects have been implicated: On the one
hand, the apparent rise in incidences below $50$ years may be
attributed to life-style
changes~\cite{SieDeSJem2014,SieFedAndMilMaRosJem2017} an effect not
incorporated in our biological model. On the other hand, recent
results support the presence of a large undetected preclinical case
burden $<50$ years, which is not reflected in the rates of colorectal
cancer observed in the SEER registries~\cite{AbuZhoAhnQinXiaKar2020}.
The latter is consistent with the fact, that people younger than $55$
are more likely to be diagnosed with late-stage disease, largely due
to sometimes for years delayed follow-up of
symptoms~\cite{SieFedAndMilMaRosJem2017}. This implies that cancer
incidences are either missed at younger ages or wrongfully assigned to
older age groups, whose rates they barely impact as the age-specific
rates increase significantly with age. Since the net impact of these
two opposing effects is unknown, we do not adjust incidence rates
below $55$ years, similar to Ref.~\cite{LueCurJeoHaz2013}. Note that
external factors, like life-style changes and screening effects, can
be indirectly captured by adjustments of the observed incidence rates
for age, birth cohort, and time period effects, for instance via
age-period-cohort (APC) epidemiological
models~\cite{LueMoo2002,MooMezTur2009,MezJeoRenLue2010,LueCurJeoHaz2013,BroEisMez2016}.
However, while such statistical adjustments compensate for secular
trends, they do not allow to determine what the incidence rates would
be in the absence of these external factors. Consequently, such
adjustments are often not invoked when comparing incidence rates to
biological models of cancer development, in particular for colon
cancer~\cite{CalTavShi2004,KimCalTavShi2004,LitVinLi2008,SotBro2012,LanKuiMisBee2020,PatCleBoz2020}.

Note that incidences of colorectal cancer differ between
sexes~\cite{LueMoo2002,LitVinLi2008,MezJeoMooLue2008,MezJeoRenLue2010,LueCurJeoHaz2013,RhyOhKimKimRhyHon2021},
races~\cite{LueMoo2002} as well as left-(distal) and
right(proximal)-sided cancers~\cite{MezJeoRenLue2010,Levetal2018}.
Since we focus with our model on the net effect of basic biological
processes and thus do not incorporate mechanisms accounting for these
differences, e.g., screening differences which account for almost half
of the disparity between colorectal incidences in white and black
populations~\cite{SieDeSJem2014}, we consider the total number of
incidences regardless of sex, race or site, as previous model
approaches~\cite{CalTavShi2004,LanKuiMisBee2020,PatCleBoz2020}.

\subsection*{Pretumor progression model based on niche competition}

In order to test the hypothesis, that the fate of tumor development is
determined by niche competition in the early phase of tumor
development, we predict the incidence rates of colon cancer by a
mathematical model based on the tumor initiation of the
adenoma-adenocarcinoma sequence. The tumor-originating cells of colon
cancer are known to be multipotential stem cells located in a niche at
the base of intestinal crypts, also called intestinal glands, which
are finger-like invaginations on the surface of the
intestine~\cite{HumWri2008,ZeuTodStaDeM2014,Sotetal2015}. On average
such a stem cell divides asymmetrically, resulting in one daughter
cell which remains as stem cell in the niche and another daughter
cell, which is committed to differentiation and leaves the crypt, such
that the number of stem cells in the niche is always roughly
conserved. However, the progeny of one stem cell can effectively
replace the progeny of another whenever symmetric divisions occur,
that is one stem cell divides into two stem cells while another
divides into two differentiated cells, which leave the crypt. During
division stem cells can mutate and thus gradually accumulate
alterations which turn them into tumor cells. Tumor cells compete with
the original stem cells via replacement in the niche and thus may at
some point dominate the entire niche (niche succession or clonal
conversion). The so fixated tumor cells may then grow across the
intestinal epithelium by crypt fission and clonal expansion of tumor
cells from the converted crypt and consequently a tumor develops. If
tumor cells are benign at that point, a premalignant lesion (adenoma)
develops, which may advance to cancer (adenocarcinoma) given a tumor
cell progresses to a malignant type (sequential progression). If tumor
cells are already malignant during fixation in the niche, an
adenocarcinoma grows directly (tunneling progression).

\begin{figure}[t!]
  \centering
  \includegraphics[scale=0.6]{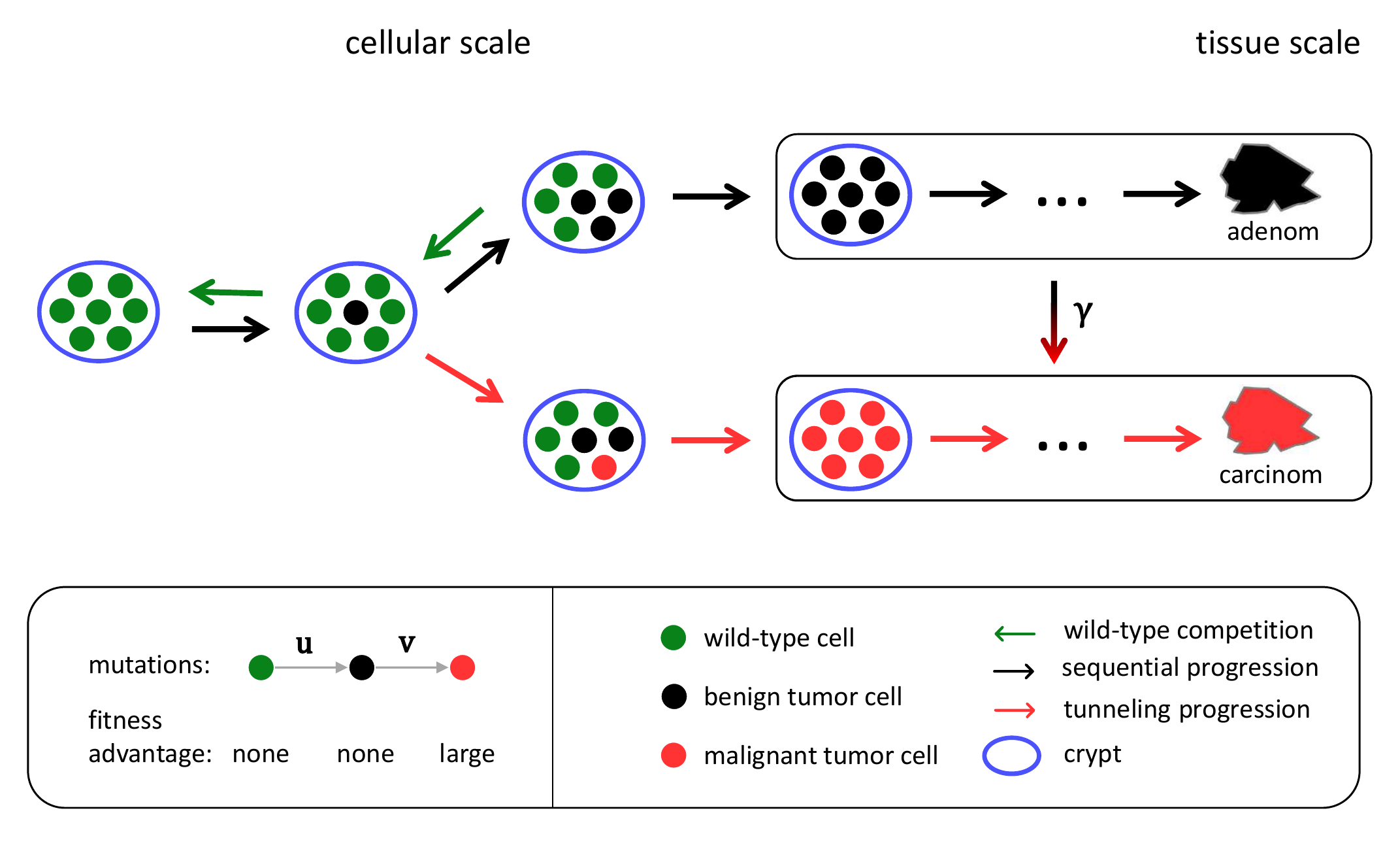}
  \caption{\label{fig:model} Illustration of the dynamics of the
    niche-competition-based pretumor progression model. Wild-type
    cells within the niche of a colonic crypt can progress to benign
    tumor cells, which can further progress to malignant tumor cells,
    during proliferation with mutation probabilities $\mutb, \mutm$.
    While wild-type and benign tumor cells neutrally compete by
    replacing each other in the niche with rate $\rr$, malignant tumor
    cells rapidly dominate the niche. We assume that once no more
    wild-type cells are present in the niche (fixation or clonal
    inversion) tumor cells establish within the tissue via clonal
    expansion. Consequently, there are two possible progression
    pathways: (i) The niche consists of malignant tumor cells at
    fixation and an adenocarcinoma grows from the niche (tunneling
    progression). (ii) The niche consists of benign tumor cells at
    fixation and the resulting tumor progresses with probability
    $\pprog$ at some point during growth to an adenocarcinoma
    (sequential progression, precancerous lesion).}
\end{figure}

We emulate this multistep process using a cell-based Moran model
representing the competition between wild-type stem cells and tumor
cells in the niche~\cite{BudDeuKliVos2015,BudDeuKliVos2019}, see
\figref{model} for illustration and \siref{model} for details: We
consider a fixed number $N$ of cells in the niche, all of which are
initially wild-type stem cells. During proliferation a wild-type cell
can mutate with probability $\mutb$ to a benign tumor cell, which in
turn can mutate with probability $\mutm$ to a malignant tumor cell.
Wild-type cells and benign tumor cells compete by replacing each other
in the niche with rate $\rr$. Eventually, via this competition,
existing benign tumor cells either go extinct or they replace every
wild-type cell in the niche. In the latter case, the tumor cells are
considered to spread and develop a clinically detectable adenoma. From
the Moran model corresponding to these dynamics, we obtain the adenoma
probability $\Pb(t)$ that an adenoma has developed from the crypt
until age $t$, see \eqref{transition-N}. Alternatively, during
competition in the niche, any benign tumor cell can acquire
malignancy. Due to the proliferative fitness advantage thus acquired,
a malignant tumor cell rapidly dominates the niche and consequently an
adenocarcinoma develops directly. From the Moran model, we obtain the
carcinoma probability $\Pm(t)$ that an adenocarcinoma has developed
from the crypt until age $t$, see \eqref{transition-E}.

Thus, there are two possible origins of colon cancer in the model: (i)
Firstly, tunneling progression, which means an adenocarcinoma develops
directly from a single niche with probability $\Pm(t)$. (ii) Secondly,
an adenoma develops in a single niche with probability $\Pb(t)$ and
progresses with probability $\pprog$ to an adenocarcinoma during
clonal expansion. Therefore, the probability that cancer has developed
from the crypt until age $t$ via this sequential progression is
$\pprog\Pb(t)$. Consequently, the combined probability that an
adenocarcinoma develops from a single crypt until age $t$, either with
benign precursor stage or directly, is $\Pm(t)+\pprog \Pb(t)$.
Inversely, the probability that a crypt has not given rise to a
cancerous tumor until age $t$ is $1-\Pm(t)-\pprog \Pb(t)$. Taking into
account that there exist $\C$ crypts in the colon and a single one
suffices as origin of a tumor, the probability $S(t)$ that a human has
not contracted colon cancer until age $t$ is
\begin{equation}
\label{eq:survival}
  S(t) = (1-\Pm(t)-\pprog \Pb(t))^\C \ , \qquad S(0) = 1 \ .
\end{equation}
From this survival function $S(t)$, the model's age-specific incidence
rate $\R(t)$, or hazard function, is computed
\begin{equation}
  \label{eq:rates}
  \begin{aligned}
  \frac{\mathrm{d}}{\mathrm{d}t}S(t) &= - \R(t) S(t) \\
  R(t) &=
  - \frac{\mathrm{d}}{\mathrm{d}t} \ln S(t)
  \end{aligned}
\end{equation}
The six parameters $N$, $\C$, $\mutb$, $\mutm$, $\rr$ and $\pprog$ of
the model are all directly measurable, independently of the emerging
age-specific incidence rates, and their values have already been
determined or estimated in previous experiments, see
\tabref{parameters}. Note that a simpler version of the model has
already been used to estimate the competition range for several types
of solid tumors from the fraction of observed benign
tumors~\cite{BudDeuKliVos2019} and to evaluate the spontaneous tumor
regression in pilocytic astrocytoma~\cite{BudDeuKliVos2015}.

The model employs some reasonable simplifications: While colon cancer
increases malignancy gradually in seven steps~\cite{VogKin2002}, we
only regard the last benign alteration and the first malignant
alteration. In particular for colon cancer, more than two steps or
more than one type of genomic instability are not expected to enhance
the agreement between a progression model and the epidemiological
data~\cite{LitVinLi2008}. Furthermore, we do not assume a particular
spatial arrangement of the stem cells in the niche, but rather use two
limiting cases of either space-free (all-to-all competition) or
one-dimensional cell arrangement (competition only with two
neighbors). Since the model's probabilities $\Pb(t)$ and $\Pm(t)$
depend monotonously on the number of cells which can compete with each
other~\cite{DurMos2015}, the incidence rates for any cell arrangement
should lie in between these two limiting cases. Note that typically
neighboring replacement is assumed, consistent with observations from
genetic lineage tracing in mouse and recent observations of epidermal
stem cells~\cite{Mesetal2018}. However, for the small number $N$ of
stem cells in a colonic crypt these two limiting cases yield almost
matching results. Furthermore, while the model allows explicitly for
two different mutation probabilities $\mutb, \mutm$, we use the same
parameter range for both, see \tabref{parameters}, although the benign
mutation may allow the malignant mutation to occur more easily
$\mutm>\mutb$. Finally, the model neglects the growth kinetics after
tumor cells fixate in a niche. The details of this growth may be very
complex and the length of the adenoma-adenocarcinoma interval depends
on size, morphology and pathological type of the
adenoma~\cite{PagRadRepHas2015} but the kinetics have been estimated
by models with clonal expansion~\cite{LueCurJeoHaz2013}. Note that a
single converted crypt does not directly lead to a carcinoma but
rather additional processes like crypt fission and evolutionary and
adaptive processes that establish a tumor micro-environment and a
permissive immune-ecology are critical for the spread and fixation of
the mutation in the tissue. We take these processes only implicitly
into account by using the replacement rate $\rr$ in the model as an
effective parameter setting the time scale. Thus, this effective rate
is expected to be smaller than the actual replacement rate, reflecting
the additional time to fixation in the tissue and diagnosis. This
smaller effective replacement rate also allows to compensate for the
simplified competition after occurrence of a single malignant tumor
cell within the niche. This is again reasonable due to the small
number $N$ of cells a malignant cell has to outcompete in a colonic
crypt. Finally, the model, while conceptually very different, produces
age-specific incidence rates with features typically known from
multistage clonal expansion models (MSCE)~\cite{MezJeoRenLue2010,
  MezJeoMooLue2008,BroMezEis2017}, that is incidence rates exhibit a
power law at young age and transition with increasing age to an
exponential increase, followed by a linear increase until
asymptotically becoming constant, see \siref{model} for details. Note
that MSCE models include an early phase without
expansion~\cite{LueCurJeoHaz2013}, resulting in a non-zero probability
that cancer arises within a single crypt or from a very small number
of initiated cells, which corresponds effectively to a fixation before
proliferation of crypts.

\section*{Results}

\begin{table*}
  \centering
  \caption{\label{tab:parameters} Table of previously reported ranges
    of the parameters used in the niche model. The upper and lower
    boundaries as well as the expected value of each parameter $\C$,
    $\mutm$, $\mutb$, or $\pprog$ are used for the parameter
    variation, see gray lines in \figref{incidence}, by which the
    effect of the parameter range on the prediction of the model is
    estimated.}
  \begin{tabular*}{1.\hsize}{p{0.3\hsize} P{0.05\hsize}
    p{0.2\hsize}| P{0.3\hsize}}
    Parameter & & & previously reported range \\
    \hline
    niche size & $N$ & & $5 \ldots 7 \ldots 15$~\cite{Nicetal2018,Baketal2014,Gabetal2022} \\
    number of crypts & $\C$ & $\left[10^7\right]$ & $1 \ldots 1.5\ldots 2$~\cite{TomVog2015,HouDesDesBerVel2002}
    \\
    mutation probabilities & $\mutb, \mutm$ & $\left[10^{-6} \right]$
                  & $1.75 \ldots 4.4 \ldots
                    7.13$~\cite{Nicetal2018}
    \\
    fraction of adenomas progressing to carcinomas & $\pprog$ &
                                                                $\left[
                                                                \%
                                                                \right]$
                  & $1.5 \ldots 5 \ldots 9.4$~\cite{GraVar2008,Steetal2013}
    \\
    stem cell replacement rate & $\rr$ & $\left[\text{
                                         y}^{-1}\text{stem cell}^{-1}\right]$
                  & $0.3 \ldots 1 \ldots 2.0$~\cite{SieMarWooTavShi2009,Nicetal2018,Gabetal2022}\\
    \end{tabular*}
\end{table*}

We apply our pretumor progression model to age-specific incidence
rates of colon cancer, see \figref{incidence}, adjusted for effects of
colorectal screening, see \siref{adjust}. We set the model's
parameters as reported for the colon in the literature, see expected
values in \tabref{parameters}, and calibrated the effective
replacement rate $\rr$ by the epidemiological data, see \siref{fit}.
The resulting prediction of the model agrees with the age-specific
incidence rate of the most recent SEER data not only in the regime of
older ages, see rates on linear scale in inset of \figref{incidence},
but also down to young ages, where incidences are more than two orders
of magnitude smaller, see rates on logarithmic scale in the main panel
of \figref{incidence}. The only exception is the single incidence rate
at $5-9$ years, for which the prediction is about four times higher
than the data. This discrepancy may result from the mentioned issue of
delayed diagnosis, which may have a particular impact at this young
age and small incidence ($5$-times smaller than incidence in next age
group $10-14$).

The robustness of the model prediction within the parameter ranges
taken from the literature is checked by a coarse parameter variation:
For all possible parameters sets, for which each of the parameters
$\C$, $\mutb$, $\mutm$, $\pprog$, either assumes the lower or upper
limit or expected value as reported in the literature, see
\tabref{parameters}, and for which $N$ assumes any integer $5-15$, the
effective replacement rate $\rr$ is fitted to the epidemiological data
resulting in an ensemble of model predictions. For illustration, this
ensemble is displayed by gray lines in \figref{incidence}, where for
clarity the opacity of each line scales inversely with the
goodness-of-fit. It turns out, that fixing any of the parameters $\C$,
$\mutm$, $\mutb$, or $\pprog$ to one of its three values while varying
all other parameters, leads to an ensemble of model predictions whose
average effective replacement rate, life-time fraction of benign tumor
and goodness-of-fit is virtually the same in either case. In contrast,
effective replacement rate and goodness-of-fit strongly depend on the
niche size $N$. In the following, we consider the parameter sets whose
goodness-of-fit is less than hundred times bigger than the smallest
occurring value, which corresponds roughly to the gray lines visible
in \figref{incidence}. Within this ensemble, the niche sizes are
predominantly $N=6-10$ with an average $N = 8 \pm 1.8$, meaning that
the model fits better for the average or lower limit of the niche
size. The fact that the niche size $N$ has such an impact on the
predictions but not the number of crypts $\C$, which also affects the
total number of stem cells, highlights the crucial role of the
competition in the niche, which is only affected by $N$ but not $\C$.

Furthermore, we estimate from the model the life-time fraction of
adenocarcinoma, which arise from colonic crypts, as
$\pprog\Pb(T) / (\Pm(T)+\pprog\Pb(T)) = 96.4 \pm 1.3 \%$ ($T=85$~y)
and the life-time fraction of benign tumor as
$\Pb(T)/(\Pm(T)+\Pb(T)) = 99.7 \pm 0.1 \%$ in agreement with the
corresponding clinical estimates of $>95\%$ and
$99\%$~\cite{DriRid1994}.

In the main panel of \figref{incidence}, the incidence rates from the
model are additionally decomposed into the contribution from (i)
tunneling progression without precancerous lesions
$-\mathrm{d}/\mathrm{d}t \ln (1-\Pm(t))^\C$ and (ii) sequential
progression $-\mathrm{d}/\mathrm{d}t \ln (1-\pprog\Pb(t))^\C$. The
incidences at younger age are dominated by the former type, while they
result almost solely from the latter one for ages beyond $40$.
Furthermore, the incidence rates from the tunneling progression
exhibit a linear slope $\sim t$ while the sequential progression leads
to a much steeper course, which resembles a Weibull distribution.
These features of the model's dynamics are consistent with the fact
that early onset gastrointestinal cancers commonly arise without
precancerous lesions while adenomatous lesions are more common in
older patients and that carcinogenesis accelerates after the age of
$40$~\cite{BedFaiBouPiaCauHil1992,MeiMorMieBaySto2001,RhyOhKimKimRhyHon2021}.
Note that the probability $\pprog\Pb(t)$ should be understood as an
upper limit for sequential progression from a single crypt as some
tumors may turn malignant during tumor growth or go extinct before the
adenoma becomes detectable. Consequently, the probability $\Pm(t)$ is
a lower limit of the fraction of tunneling progression. This implies
that the age at which the contributions of both progressions match
$\Pm(t)=\pprog\Pb(t)$ is underestimated by the model in
\figref{incidence}.

The ensemble of model predictions covers a range of effective
replacement rates $0.01-0.06 \text{ y}^{-1}$ per stem cell with an
average $\rr = 0.024 \pm 0.01 \text{ y}^{-1}$ per stem cell. While an
effective rate smaller than the actual rate is in principle expected
due to the neglected timescale between fixation in the crypt and
clinical detection, this effective rate is considerably (factor
$5-200$) smaller than the physiological replacement rates reported in
the literature, see
\tabref{parameters}~\cite{Nicetal2018,Gabetal2022}. This suggests
that, while the course and composition of the age-specific incidence
rates seems to be determined by the competition in the crypt, the time
scale is predominantly set by the processes succeeding the conversion
of crypts. Note that the physiological replacement rate itself has
been recently updated~\cite{Nicetal2018,Gabetal2022} by two orders of
magnitude compared to previous estimates, which puts the replacement
rate close to the cell division
rate~\cite{LopKleSimWin2010,Baketal2014}. Finally, note that we
adjusted the epidemiological incidence rates for effects of colorectal
screening via different estimates and observe similar correspondence
between model and data in all cases, see \siref{adjust}.

\begin{figure}[ht!]
  \centering
\hspace*{-1cm}\includegraphics{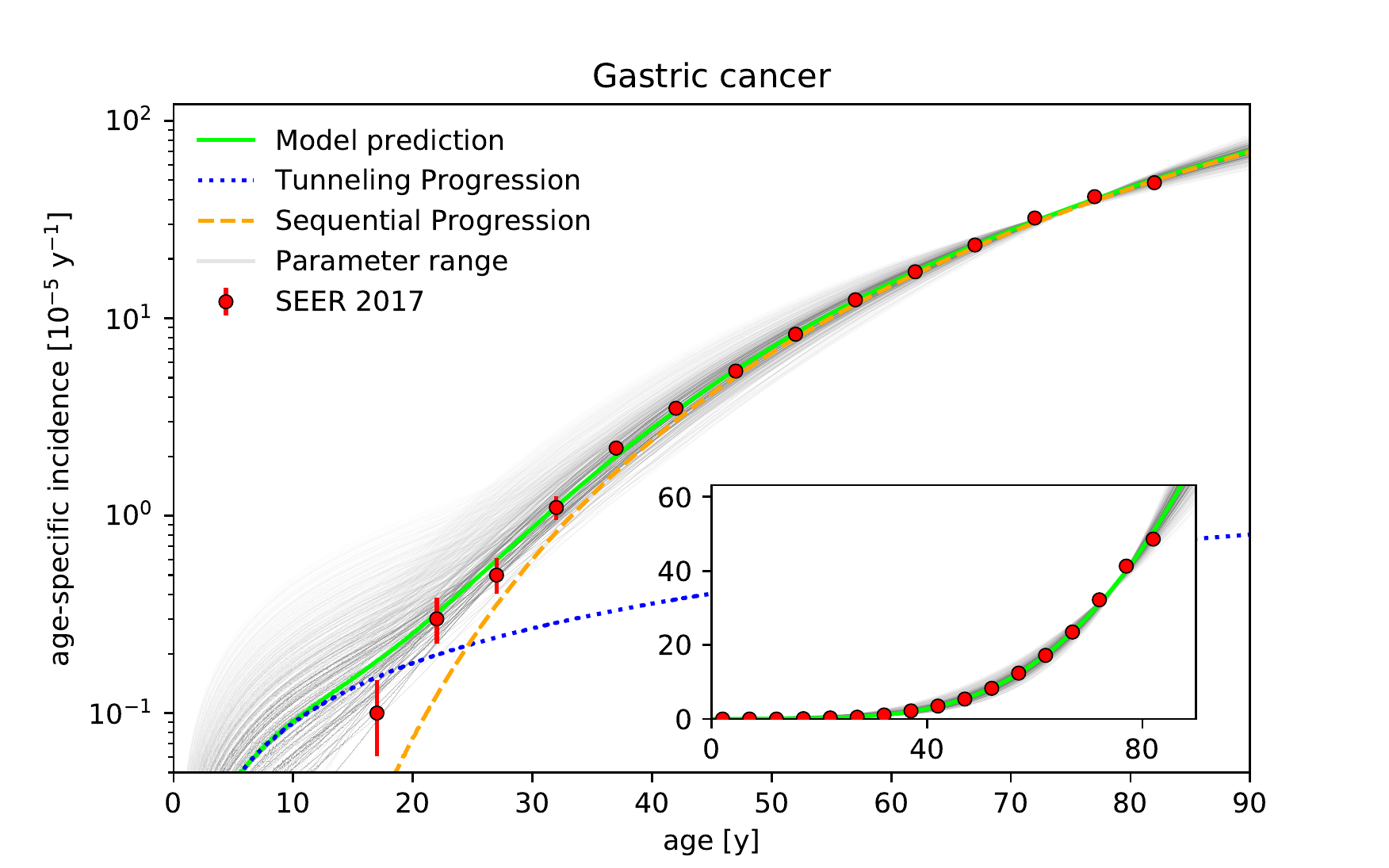}
\caption{\label{fig:gastric-cancer} \textbf{Niche competition could
    model age-specific incidence rates of gastric cancer.}
  Age-specific incidence rates of gastric cancer predicted by the
  model show correspondence to epidemiological data (SEER
  database~\cite{SEER2017}, displayed curves and points analogous to
  \figref{incidence}). Since parameters are not available as detailed
  as for the colon, the range of the parameters $N$, $\mutb$, $\mutm$,
  and $\pprog$ are assumed to match the ranges for the colon, see
  \tabref{parameters}, while the range of the number of gastric glands
  $\C = 4 - 16 \cdot10^6$ is estimated as described in the text.
  Exemplary parameter set with $N = 7$, $\C = 8\cdot10^6$,
  $\mutb=4.4\cdot10^{-6}$, $\mutm = 1.75\cdot10^{-6}$, and
  $\pprog = 5\%$ is highlighted in green. The effective replacement
  rate $\rr$ is found to be considerably smaller than replacement
  rates of colonic crypts ($\rr = 0.014 \text{ y}^{-1}$ per stem cell
  for the green curve and $\rr = 0.02 \pm 0.01\text{ y}^{-1}$ for the
  gray curves illustrating the parameter range).
}
\end{figure}

The model is additionally applied to gastric cancer, whose
tumor-originating cells are proposed to be also compartmentalized into
niches, see \figref{gastric-cancer}. In the context of gastric cancer,
it is known that gastric corpus and antrum have distinct stem cells
regarded as the tumor-originating cells, which are also structured in
stem cell niches within gastric glands~\cite{HayFoxWan2017}. Since the
corresponding parameters are not available as detailed as for the
colon, the range of the parameters $N$, $\mutb$, $\mutm$, and $\pprog$
are assumed to match the ranges for the colon, see
\tabref{parameters}, while the number of gastric glands
$\C = 4 - 16 \cdot10^6$ is estimated from the approximate surface of
the stomach $\sim 800$~cm$^2$ and the density of glands
$135 \text{ mm}^{-2}$~\cite{KurKurDmiVce2001}. We find a good
agreement between model prediction and epidemiological data, see
\figref{gastric-cancer}. The obtained effective replacement rate
$\rr = 0.02 \pm 0.01 \text{ y}^{-1}\text{stem cell}^{-1}$ is similar to
the value obtained from the model for the colon, see
\figref{incidence}.

\begin{figure}[ht!]
  \centering
\hspace*{-1cm}\includegraphics{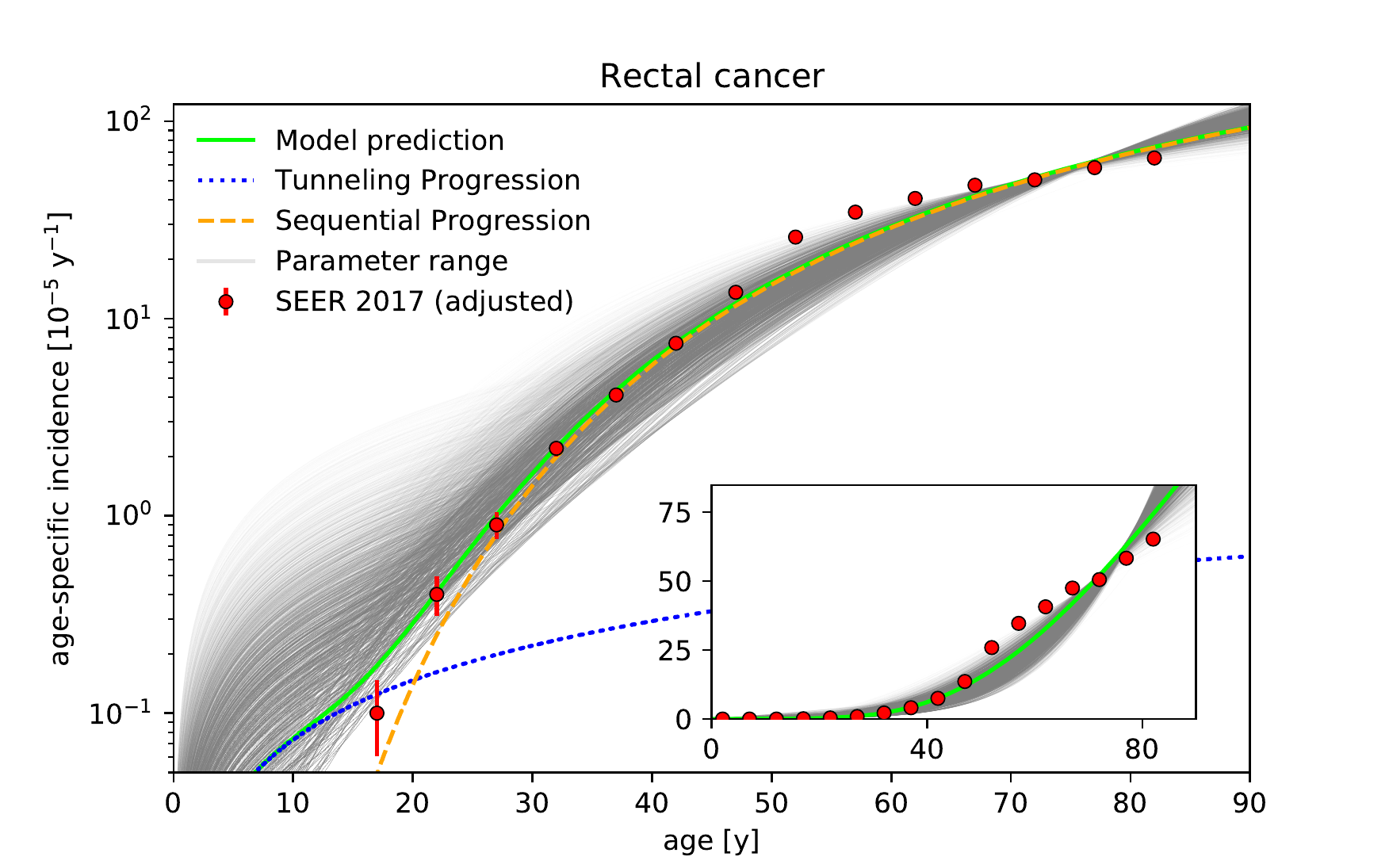}
\caption{\label{fig:rectal-cancer} \textbf{Niche competition model
    corresponds well to age-specific incidence rates of rectal cancer
    for age groups below $50$.} Age-specific incidence rates of rectal
  cancer predicted by the model shows rough correspondence with
  epidemiological data (SEER database~\cite{SEER2017}, displayed
  curves and points analogous to \figref{incidence}). Same parameters
  as for the the colon are assumed, see \tabref{parameters}, except
  for the number of crypts $\C$ which is assumed to be ten times
  smaller $\C = 10^6 - 2\cdot10^6$ reflecting that the rectum has on
  average a tenth of the length of the colon. Exemplary parameter set
  with $N = 8$, $\C = 10^6$, $\mutb=1.75\cdot10^{-6}$,
  $\mutm = 1.75\cdot10^{-6}$, and $\pprog = 5\%$ is highlighted in
  green. The effective replacement rate $\rr$ is found to be
  considerably smaller than replacement rates of colonic crypts
  ($\rr = 0.05 \text{ y}^{-1}$ per stem cell for the green curve and
  $\rr = 0.04 \pm 0.02\text{ y}^{-1}$ for the gray curves illustrating
  the parameter range).}
\end{figure}

The model is additionally applied to rectal cancer, whose
tumor-originating cells are also compartmentalized into niches, see
\figref{rectal-cancer}. For rectal cancer, it is reasonable to assume
the same parameters as for the colon, see \tabref{parameters}, except
for the number of crypts $\C$. Since the rectum is on average one
order of magnitude shorter than the colon, the number of niches $\C$
is assumed to be ten times smaller $\C = 10^6 - 2\cdot10^6$. The
resulting prediction of the model displays a good qualitative
correspondence for the age-groups below $45$, see
\figref{rectal-cancer}. However, there are considerable deviations in
the age groups above $50$ years. Still, the model predictions displays
a rough visual correspondence to the data, capturing essential,
qualitative characteristics, which is not self-evident considering
that only a single parameter has been fitted. In addition, the data
points at older age groups may be problematic due to an underestimated
impact of rectal screening. Risk reduction is known to vary by subsite
of the colon and rectum~\cite{BreAltStoHof2015} while the estimates of
this risk reduction are for colorectal
cancer~\cite{KahImpJulRex2009,BreAltStoHof2015,Levetal2018}, whose
incidences are dominated by incidences of colon cancer. Indeed,
assuming higher reduction of annual incidences for rectal cancer due
to screening leads to substantially better correspondence between
model and data, see \figref{rectum-inc} and \figref{rectum-max}. In
any case, according to our model prediction the incidences of the
rectum are dominated by sequential progression even at small ages. The
obtained effective replacement rate
$\rr = 0.04 \pm 0.02 \text{ y}^{-1}$ per stem cell is again
considerable smaller than the range reported for the colon. In
particular, the obtained effective replacement rates $\rr$ for both
rectal and gastric cancer are similar to the one obtained for colon
cancer. However, compared to colon cancer, contribution of tunneling
progression to the incidence rates of rectal and gastric cancer result
are much smaller with deviations from the sequential progression only
visible below $25$ years.

\section*{Discussion}

We employed a Moran model representing the pretumor competition
between wild-type and tumor stem cells in colonic crypts to
quantitatively reproduce age-specific incidences of colon cancer.
Additionally, the model predicts the fraction of incidences
corresponding to either (i) adenocarcinoma or (ii) benign polyps,
which progressed to adenocarcinoma, and we find these fractions in
agreement with common clinical estimates. Furthermore, the
age-dependency of these fractions is consistent with the occurrence of
precancerous lesions in different age regimes. In particular, this
highlights the importance of cancer screening after the age of $40$.
Note that all growth and clonal expansion processes after the fixation
of tumor cells in the tumor originating niche are not explicitly
incorporated into the model, but only implicitly contained in an
effective replacement rate. We find this effective rate to be
substantially lower than the actual physiological replacement
rate~\cite{Nicetal2018,Gabetal2022}. Thus, the agreement between model
and epidemiological data supports the notion that the fate of tumor
development may be majorly determined by the early phase of tumor
development, while the time scale of tumor development is primarily
set by processes of clonal expansion and tumor growth.

We emphasize that our model does not suggest, that a single converted
crypt results inevitably into a macroscopic carcinoma, which would be
biologically untenable. Rather all processes between conversion of
crypts and diagnosis of a macroscopic malignancy are omitted, since
the purpose of the model is to estimate the impact of the earliest
phase of tumor development and the architecture of the tissue
containing the tumor originating cells. This does not mean that the
omitted processes are negligible. Instead, the relatively small
effective replacement rate obtained implies that the processes
succeeding the clonal conversion of crypts set the time scale of the
incidence rates. However, the results suggest that the shape and
composition of the incidence rates may be determined by the structure
of the tissue hosting the tumor-originating cells and the competition
dynamics inside the crypts.

Note that in contrast to most previous pretumor models, which usually
introduce four or more parameters that are calibrated using the
epidemiological age-specific cancer incidences, both for
colon~\cite{LitVinLi2008,MezJeoRenLue2010,SotBro2012,LanKuiMisBee2020}
or other types of
cancer~\cite{MdzShe2010,Bro2011,LueCurJeoHaz2013,BroEisMez2016,RhyOhKimKimRhyHon2021,MezJeoMooLue2008,GroBraHolHauKunTreAalMog2011},
our model is solely based on directly measurable parameters, which are
already known from clinical and biological studies on the colon
independent of the epidemiological incidence rates. Consequently, the
model only requires a single fit parameter, the effective replacement
rate, which both sets the time scale of the niche-competition as well
as compensates for the neglected time-span between fixation in the
niche and clinical detection.

The model is additionally applied to rectal and gastric cancer, whose
tumor-originating cells are also proposed to be compartmentalized into
niches, but exhibit incidence rates considerably smaller than for
colon cancer. For this we only adapt the number of niches while
assuming the same ranges as for the colon for all other parameters.
Despite this simplification, we observe reasonable correspondence
between predicted and epidemiological incidences rates. Furthermore,
the obtained effective replacement rates are similar for all three
cases. In combination, this points towards similar dynamics of tumor
development for colon, rectum and stomach.

Note that for colorectal cancer, two-step models have been considered
before to capture the dynamics of adenomatous and malignant tumor
cells~\cite{CalTavShi2004,KimCalTavShi2004,LitVinLi2008,LanKuiMisBee2020}.
Furthermore, the distribution of adenoma and adenomacarcinoma sizes at
different ages has been inferred from a model~\cite{LanKuiMisBee2020},
clonal expansion dynamics of colorectal cancer has been derived from
intra-tumor heterogeneity~\cite{SieMarWooTavShi2009} and even the
competition in colonic crypts has been
discussed~\cite{CalTavShi2004,KimCalTavShi2004}. In contrast to these
models, we explicitly incorporate competition of benign tumor cells
with wild-type cells taking into account their spatial structure,
while neglecting other processes considered in these models. The
agreement of the prediction of our model and the epidemiological
incidence rates emphasizes the dominant role that competition in
niches may play in tumor development. This is interesting in the
context of previous investigations of colorectal and gastric cancer
using a multistage clonal expansion (MSCE) model, which suggest that
the form of the age-specific incidence rates is determined by the
dynamics until the fixation of tumor cells, while the growth dynamics
after fixation only cause a slight time shift~\cite{LueCurJeoHaz2013}.
Future approaches could incorporate the model of crypt competition as
process of tumor initiation into the MSCE model and examine how this
affects the parameters of the model and their robustness.

Our model does not discern between patient subgroups such as sex,
race, and cancer site. In principle, the model could be fitted to each
of these subgroups, but note that there are considerable variations in
the level of reporting across them, e.g., screening differences
account for almost half of the disparity between colorectal incidences
in white and black populations~\cite{SieDeSJem2014}. Moreover, the
model assumes that all parameters are shared across the population,
because the goal is not to predict individual cancer risks. For this,
one would have to take into account individual predispositions
regarding for instance genetics, immune system or lifestyle.

While the model is inspired by colonic crypts, it is potentially
relevant for tumor development in a wide range of tissues. Firstly,
the model can be applied to cancer types with similar spatial
structure of the tumor-originating cells, as in the case of rectal or
gastric cancer~\cite{HayFoxWan2017}. Secondly, the model may also
apply to types of cancer whose tumor-originating cells are not
explicitly compartmentalized into niches. The Moran model within the
niche has already been used to estimate an upper limit for the number
$N$ of cells competing with each other from the fraction of clinically
observed benign tumors for several types of
cancer~\cite{BudDeuKliVos2019}. This so called competition range $N$
is surprisingly small for a wide range of tumor types, less than
$3000$ cells, even for tumors in tissue without explicit stem cell
compartments like the crypts. Note that for a large competition range
$N$ the competition between the malignant tumor cells and the other
cells in the niche may not be negligible anymore, but increase the
time scale until clonal conversion in the niche. Thus, the model has
to be extented in the future by this competition to be applicable to
types of cancer with large competition ranges, such as hepatocellular
carcinoma~\cite{OhgKle2013} or glioblastoma~\cite{VijElaKha2015}.

Application of the model of niche competition to cancer in tissues
other than the colon is reasonable for two more reasons: Firstly, most
other solid tumors have an age-specific incidence qualitatively very
similar to colon carcinoma~\cite{SotBro2012}. Secondly, incidence
rates resulting from the model share typical features with incidence
rates from multistage clonal expansion
models~\cite{MezJeoMooLue2008,BroMezEis2017,MezJeoRenLue2010}, which
already achieve good matches with epidemiological data for several
types of
cancer~\cite{MezJeoMooLue2008,MooMezTur2009,MezJeoRenLue2010,LueCurJeoHaz2013,MezCha2015,BroEisMez2016,BroMezEis2017}.
Note that for most tissues, much less is known about tumor initiation
and development until detection than for the colon. Applying the model
to such types of cancer may give additional insight into the hardly
observable dynamics of early cancer development.

\section*{Acknowledgments}
We acknowledge support by the EU, the European Social Fund (ESF) and
by tax funds on the basis of the budget passed by the Saxon state:
SAB-Nr. 100382145. The funding provided salary for (A.V.-B., S.L.,
D.H.). The funders had no role in study design, data collection and
analysis, decision to publish, or preparation of the manuscript.

\newpage

\renewcommand{\theequation}{S\arabic{equation}}
\setcounter{section}{1} 
\setcounter{subsection}{0} 
\renewcommand{\thesubsection}{\Alph{subsection}}
\setcounter{equation}{0}  
\renewcommand{\thefigure}{\Alph{figure}}
\setcounter{figure}{0}  
\renewcommand{\thetable}{\Alph{table}}
\setcounter{table}{0}  

\section*{Supporting Information (SI)}

\subsection{Moran Model of niche competition\label{sec:model}}

\begin{figure}[ht!]
  \centering
\includegraphics[scale=0.6]{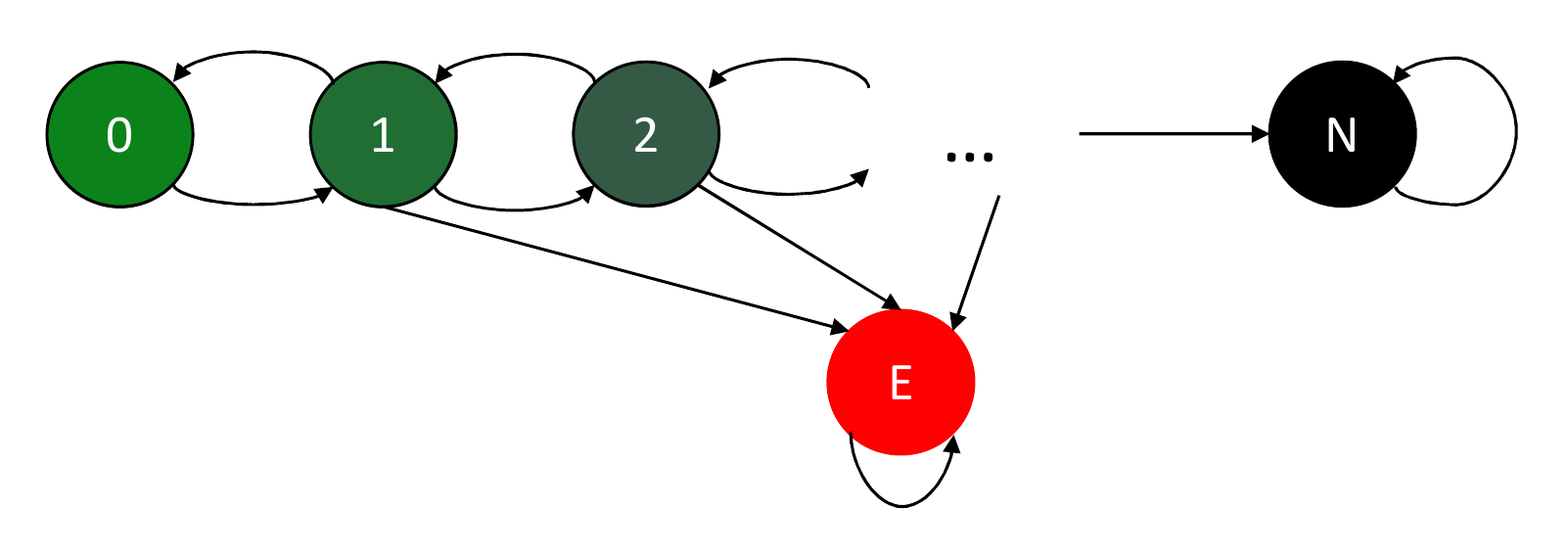}
\caption{\label{fig:markov-sketch} Sketch of the markov chain modeling
  the competition within a niche. The transition rates corresponding
  to the arrows are expressed in \eqref{all-to-all} and \eqref{ring}}
\end{figure}

We use a three-type Moran model with mutation to quantify the stem
cell dynamics in the niche of a single colonic crypt as described in
the main manuscript. The niche consists of $N$ cells, each of which
may be wild-type, benign or malignant. A cell in the niche is replaced
by the offspring of another cell with replacement rate $N\cdot\rr$,
where $\rr$ is the replacement rate per stem cell. At replacement the
offspring can additionally acquire mutations, i.e. a wild-type cells
mutates with probability $\mutb, 0\leq \mutb < 1$ into a benign tumor
cell and a benign tumor cell mutates with probability
$\mutm, 0\leq \mutm < 1$ into malignant tumor cell. Formally, the
model is a Markov process $(X_t)_{t\geq0}$ on the state space
$\S = \{0, 1, 2, \ldots, N, E \}$, see sketch in
\figref{markov-sketch} for an illustration. Here, the states
$0, \ldots, N$ correspond to the number of benign tumor cells in the
niche, while the remaining cells in the niche are wild-type stem
cells, and the state $E$ represents the presence of a single malignant
tumor cell in the niche. The dynamics is determined by a rate matrix
$Q := (q(k, l))_{k, l\in \S}\in \R^{\S\times\S}$. For the space-free
limit, i.e., when every cell is able to replace any other cell in the
niche, the rate matrix $Q_{\text{sf}}$ is given by
\begin{equation}
  \label{eq:all-to-all}
  Q_{\text{sf}} =
  \left\{
\begin{aligned}
        &q(k, k - 1)&=\ & \rr\frac{k(N-k)}{N-1}(1-\mutb) &, 1\le k \le N-1 \\
        &q(k, k + 1)&=\ & \rr\left(\frac{(N-k)(N-k-1)}{N-1}\mutb +
          \frac{k(N-k)}{N-1}(1-\mutm)\right) &, 0\le k \le N-1 \\
        &q(k, k)    &=\ & - \sum\limits_{i=0}^{N+1} q(k,i) &, 0\le
        k\le E \\
        &q(k, E)    &=\ & \rr k \mutm &, 1\le k \le N-1 \\
        &q(k, l)    &=\ & 0 &, \text{else}
    \end{aligned}\right. .
\end{equation}
The factors containing $N$ and $k$ result from
combinatorics~\cite{Kom2006,BudDeuKliVos2015,BudDeuKliVos2019}. The
other geometric limiting case is the $1D$ model in which cells are
considered to be ordered on a ring, i.e. each cell can only replace
one of its two neighbors. Analogously, the rate matrix $Q_{\text{1D}}$
for this case is given by~\cite{Kom2006,BudDeuKliVos2019}
\begin{equation}
  \label{eq:ring}
  Q_{\text{1D}} =
  \left\{
\begin{aligned}
        &q(0, 1)&=\ & \rr N \mutb &, \\
        &q(k, E)    &=\ & \rr k \mutm &, 1\le k \le N-1 \\
        &q(k, k + 1)&=\ & \rr (1-\mutm) + O(\mutb) &, 1\le k \le N-1 \\
        &q(k, k - 1)&=\ & \rr (1-\mutb) &, 1\le k \le N-1 \\
        &q(k, k)    &=\ & - \sum\limits_{i=0}^{N+1} q(k,i) &, 0\le k\le E \\
        &q(k, l)    &=\ & 0 &, \text{else}
    \end{aligned}\right. ,
\end{equation}
where, due to $\mutb \ll 1$, we neglect the term $O(\mutb)$, which
describes an additional benign mutation of a wild type cell. This
reasonable approximation implies that benign mutants only occur on a
single connected interval of the ring of cells. A real crypt will have
some geometric restrictions that lie in between these to limiting
cases $Q_{\text{1D}}$ and $Q_{\text{sf}}$.

For either rate matrix $Q$, the dynamics of the niche, i.e. the
distribution
$\mathbf{p} = \left(p_0, p_1, \ldots, p_{N-1}, p_N, p_E\right) \in [0,
1]^\S$ on the state space $\S$ at age $t$ which evolved from an initial
state $\mathbf{p}(0)$ at age $0$, is given by
\begin{equation}
\frac{\mathrm{d}}{\mathrm{d}t} \mathbf{p}(t) = \mathbf{p}(t) Q
\end{equation}
which yields the solution
\begin{equation}
  \label{eq:transition}
    \mathbf{p}(t) = \mathbf{p}(0) \exp(Q t),
\end{equation}
for $t\geq 0$, with the transition matrix $\exp(Q t)$ given by a
matrixexponential function. The initial state is set to
$\mathbf{p}(0) := (1, 0, \ldots, 0)$, i.e. the niche starts at birth
with only wild-type cells in the crypt. The probability $\Pb(t)$ that
the complete niche consists of benign tumor cells until age $t$ is
\begin{equation}
  \label{eq:transition-N}
\Pb(t) = \left(\mathbf{p}(0) \exp(Q t)\right)_N
\end{equation}
and the probability $\Pm(t)$ that the first malignant cells emerges
until age $t$ is
\begin{equation}
  \label{eq:transition-E}
\Pm(t) = \left(\mathbf{p}(0) \exp(Q t)\right)_E \ .
\end{equation}

We further list some analytic and empiric properties of the model,
which demonstrate that the incidence rates generated by the model
display features typical for multistage clonal expansion models, whose
incidence rates transition (from young to old age) from power-law to
exponential to linear to constant asymptote~\cite{MezJeoRenLue2010,
  MezJeoMooLue2008,BroMezEis2017}: Previously, an analytic expression
for the asymptotic absorbtion probabilities
$\lim\limits_{t\to\infty}\Pb(t)$ and $\lim\limits_{t\to\infty}\Pm(t)$
into state $N$ and state $E$, respectively, has been
derived~\cite{BudDeuKliVos2015,BudDeuKliVos2019}. These asymptotic
probabilities are predominantly determined by the relation of the
niche size $N$ and the probability $\mutm$ to progress from a benign
tumor cell to a malignant one. A small niche $N \sqrt{\mutm} \ll 1$
($N \sqrt[3]{\mutm} \ll 1$ for $1D$ model) is primarily absorbed in
the benign state, whereas a large niche $N \sqrt{\mutm} \gg 1$
($N \sqrt[3]{\mutm} \gg 1$ for $1D$ model) is primarily absorbed in
the malignant state $E$, while intermediate niche sizes lead to finite
probabilities for both states. This property has already been
exploited to estimate upper limits for the effective niche sizes of
several types of cancer from their recored benign tumor
fraction~\cite{BudDeuKliVos2019}. Furthermore, the probability
$\tilde{S}(t) = (1 - \Pm(t) - \Pb(t))$ to not be in either state $N$
or $E$ until time $t$ decreases exponentially
$\tilde{S}(t \gg 1) \sim \exp(\alpha t)$ for sufficiently large times
$t$, where the rate $\alpha<0$ is the largest eigenvalue of the rate
matrix that results from limiting the rate matrix $Q$ only to the
non-absorbing states $\{0, 1, 2, \ldots, N-1 \}$. Since the
probability $\tilde{S}(t)$ is qualitatively very similar to the
survival probability $S(t)$ in \eqref{survival}, one may infer that
the age-specific incidence rates $R(t)$ generated by the model
according to \eqref{rates} should become constant asymptotically
$R(t\gg1) \approx |\alpha| \C$. Moreover, the probability of tunneling
progression $\Pm(t)$, whose contribution is maximal at young ages,
follows a power law $\Pm(t)\sim t$. This is in accordance with the
argument of Armitage and Doll, that in a pure multistage model the
incidence rates follow a power law $\R(t) \sim t^{\beta -1}$, with the
number $\beta$ of mutations necessary for a cell to become
malignant~\cite{ArmDol1954,ArmDol2004}, which is $\beta = 2$ in our
model. In contrast, we observe numerically that the probability of
sequential progression $\Pb(t)$ matches the cumulative distribution
function of a Weibull distribution
$\Pb(t) = 1-\exp\left(-a t^b\right)$, which implies for barely
changing survival probabilities $S(t)\approx 1$, see \eqref{rates},
\begin{equation}
  R(t) = - \frac{\mathrm{d}}{\mathrm{d}t} \ln S(t) = -\frac{\frac{\mathrm{d}}{\mathrm{d}t} S(t)}{S(t)} \underset{S(t)\approx 1}{\approx}
  -\frac{\mathrm{d}}{\mathrm{d}t} S(t)
\end{equation}
that the contribution of the sequential progression to the incidence
rates follows a Weibull distribution. Note that it is common for
models of cancer development that the predicted incidence rates (or
hazard function) effectively represent a Weibull
distribution~\cite{SotBro2012,MdzShe2010,MdzGleKinShe2009,CalTavShi2004,GroBraHolHauKunTreAalMog2011,Bro2011}.
The assumption $S(t) \approx 1$ is reasonable, e.g., for colon cancer,
since colon cancer is overall a relatively minor cause of death and
has a low lifetime incidence.

\subsection{Fit of replacement rate \label{sec:fit}}

We determine the least known parameter, the effective replacement rate
$\rr$, by fitting the model to the epidemiological age-specific
incidence rates $\R_{\text{SEER}}(t)$. First, we compute for given set
of parameters $N, \mutb, \mutm, \C$, and $\pprog$ from the literature
the survival probability $S(\t)$ of the model in dimensionless time
$\t = t/\rr$, using \eqref{survival}, \eqref{transition-N}, and
\eqref{transition-E} up until a sufficient long time $\t_{\text{max}}$
such that the probability $1-S(\t_{\text{max}})$ is larger than the
lifetime risk of colon cancer ($\geq 4.8\%$~\cite{SEER2017}) to ensure
that the corresponding incidence rates $\R(\t)$, see \eqref{rates},
cover the range of ages of the epidemiological data.

When the predicted $\R(t)$ and epidemiological rates
$\R_{\text{SEER}}(t)$ can be matched very close at all ages, the rate
$\rr$ can be directly obtained by matching the corresponding survival
probabilities $S(t)$, . Since the probabilities $S(t)$ are
monotonously decreasing with age $t$, we can identify each data point
$(t_i, S_{\text{SEER}}(t_i))$ to a simulated point $(\t,S(\t))$ with
$S_{\text{SEER}}(t_i) = S(\t)$. Thus, each data point $i$ gives an
estimate of the replacement rate $\rr_i = \t/t_i$ and we fit the
replacement rate as their mean $\rr = <\rr_i>$. The standard deviation
$\sqrt{<(\rr_i-<\rr_i>)^2>}$ of the replacement rates may then serve
as a measure of the quality of a single fit.

However, since deviations between predicted and epidemiological rates
predicted $\R(t)$ and epidemiological rates $\R_{\text{SEER}}(t)$ may
cancel out in the corresponding survival probabilities, it is better
to fit the rates directly. We use non-linear least-square minimization
provided by the python package \emph{lmfit}~\cite{NewSteAllIng2014} to obtain the
effective rate $\rr$ by minimizing the residuum
\begin{align}
  \R_{\text{SEER}}(t_i)-\R(t_i) = \R_{\text{SEER}}(t_i) +
  \frac{\mathrm{d}}{\mathrm{d}t} \ln S(\t = \rr \cdot t_i) \qquad
  \forall t_i \ .
\end{align}
For a given parameter set $N, \C, \mutb, \mutm, \pprog$ the
goodness-of-fit $\chi^2$ is quantified by the relative square
deviation between the rescaled rate $R(t)$ of the model and the data
$\R_{\text{SEER}}(t_i)$
\begin{align}
  \chi^2 = \sum\limits_i
  \frac{\left[\R_{\text{SEER}}(t_i)-R(t_i)\right]^2}{\R_{\text{SEER}}^2(t_i)}
  \ .
\end{align}
For the variation of the parameters $N, \C, \mutb, \mutm, \pprog$ the
effective replacement rate $\rr$ is reported as average and standard
deviation of the ensemble of fits.

\subsection{Adjustments for colorectal screening \label{sec:adjust}}

\begin{figure}[ht!]
  \centering
\hspace*{-1cm}\includegraphics{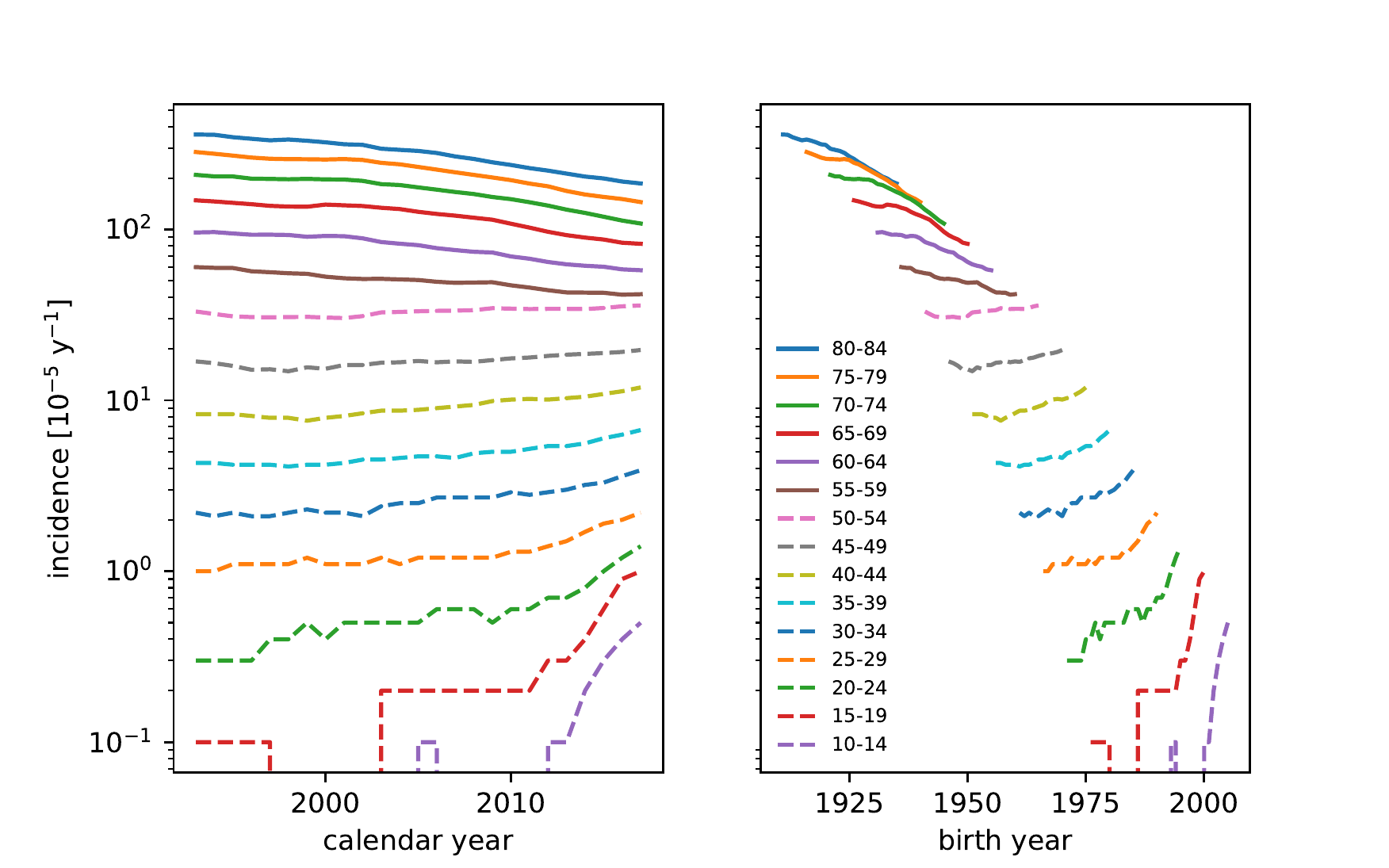}
\caption{\label{fig:trend} \textbf{Secular trends for colon cancer.}
  Incidence rates of colon cancer shown as function of calendar year
  (left panel) and birth year (right panel) based on the archive of
  the SEER database 1993-2017~\cite{SEERall}. While pronounced secular
  trends are visible, they do not affect all age groups equally.
  Rather rates are increasing for age groups below $55$ (dashed lines)
  years, while increasing above (solid lines).}
\end{figure}

The SEER data for colorectal cancer is fraught with two important
secular trends, i.e., a downward trend of incidences at ages above 55
due to widespread colorectal screening and
resection~\cite{SieDeSJem2014,Baietal2015,KahImpJulRex2009,Levetal2018,CarZhuGuoHeiHofBre2021}
and an upward trend of incidences of early onset colorectal cancer for
ages below 50, whose origin is less understood but has been attributed
to both life-style
changes~\cite{SieDeSJem2014,SieFedAndMilMaRosJem2017} as well as the
detection of prevalent subclinical
cases~\cite{SieFedAndMilMaRosJem2017,AbuZhoAhnQinXiaKar2020}. The
incidences exhibit these age-group-dependent trends both over calendar
and birth year, see \figref{trend}, especially a strong birth cohort
effect for younger ages. While these trends may be indirectly
compensated for by age-period-cohort (APC) epidemiological
models~\cite{LueMoo2002,MooMezTur2009,MezJeoRenLue2010,LueCurJeoHaz2013,BroEisMez2016},
such statistical adjustments do not allow to determine what the
incidence rates would be without these external effects. Consequently,
such adjustments are rarely invoked when comparing incidence rates to
biological models of cancer development, in particular for colon
cancer~\cite{CalTavShi2004,KimCalTavShi2004,LitVinLi2008,SotBro2012,LanKuiMisBee2020}.
One approach could be to perform the APC adjustments, properly
anchored for identifiability, on the relevant rates of the biological
model, e.g., the replacement rate $\rr$, to identify which parameters
are most sensitive to the secular trends. Our approach is to directly
incorporate independent estimates of the impact of colorectal
screening.

\begin{figure}[ht!]
  \centering
\hspace*{-1cm}\includegraphics{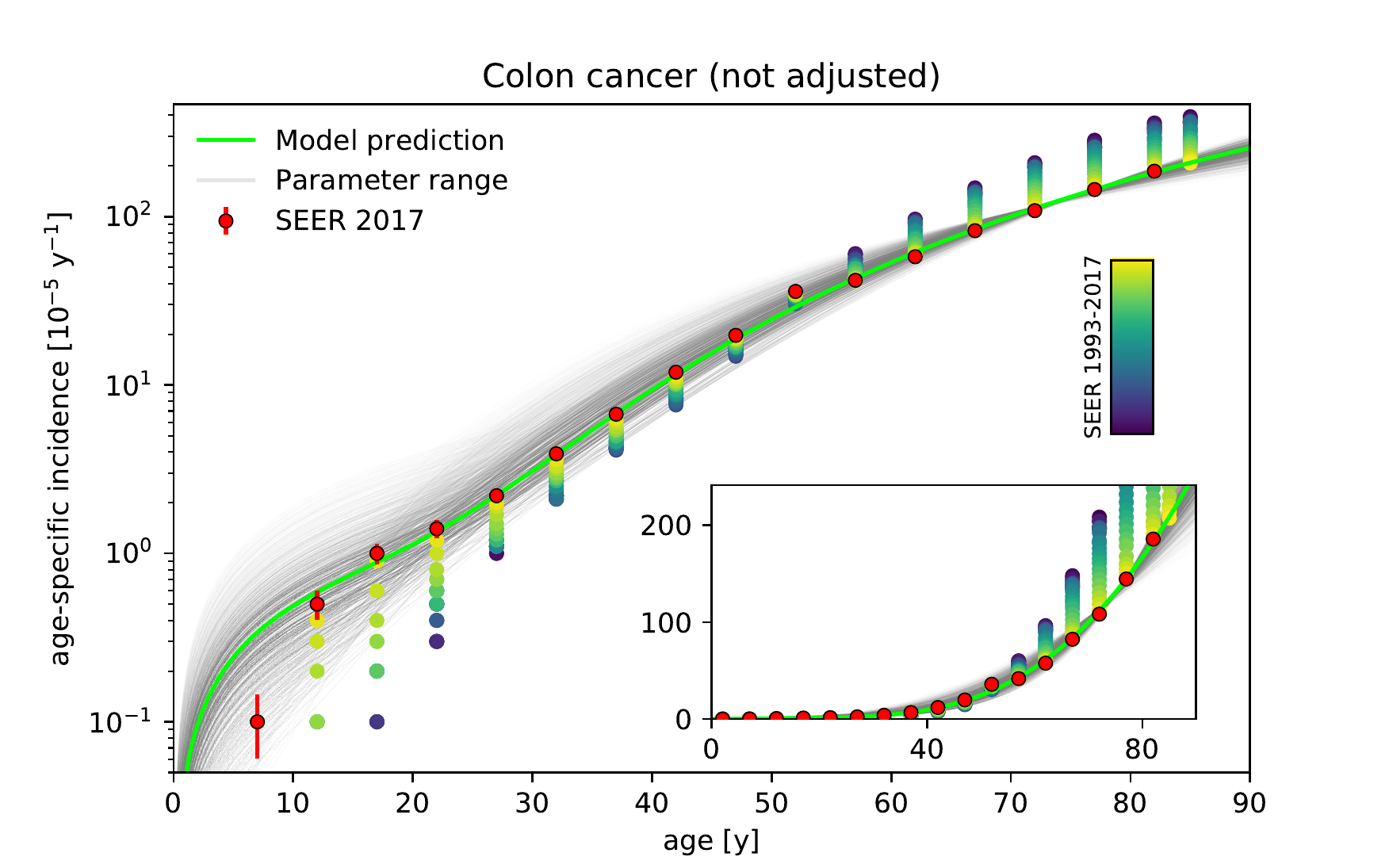}
\caption{\label{fig:colon-no-adjust} \textbf{Model comparison to colon
    data without adjustments for colorectal screening.} Age-specific
  incidence rates of colon cancer predicted by the model and
  epidemiological data (SEER database~\cite{SEER2017} unadjusted,
  displayed curves and points analogous to \figref{incidence}). As
  illustration of secular trends, the epidemiological rates from the
  archive of the annual SEER database~\cite{SEERall} are displayed by
  points colored according to the release year of the corresponding
  report from 1993 (dark blue) to 2017 (yellow). Exemplary parameter
  set with $N = 8$, $\C = 2\cdot10^7$, $\mutb=1.75\cdot10^{-6}$,
  $\mutm = 4.4\cdot10^{-6}$, and $\pprog = 9.4\%$ is highlighted in
  green. The effective replacement rate $\rr$ covers a range
  $0.01-0.06 \text{ y}^{-1}$ per stem cell with an average
  $\rr = 0.02 \pm 0.01\text{ y}^{-1}$ ($\rr = 0.02 \text{ y}^{-1}$ per
  stem cell for the green curve).}
\end{figure}

\begin{figure}[ht!]
  \centering
\hspace*{-1cm}\includegraphics{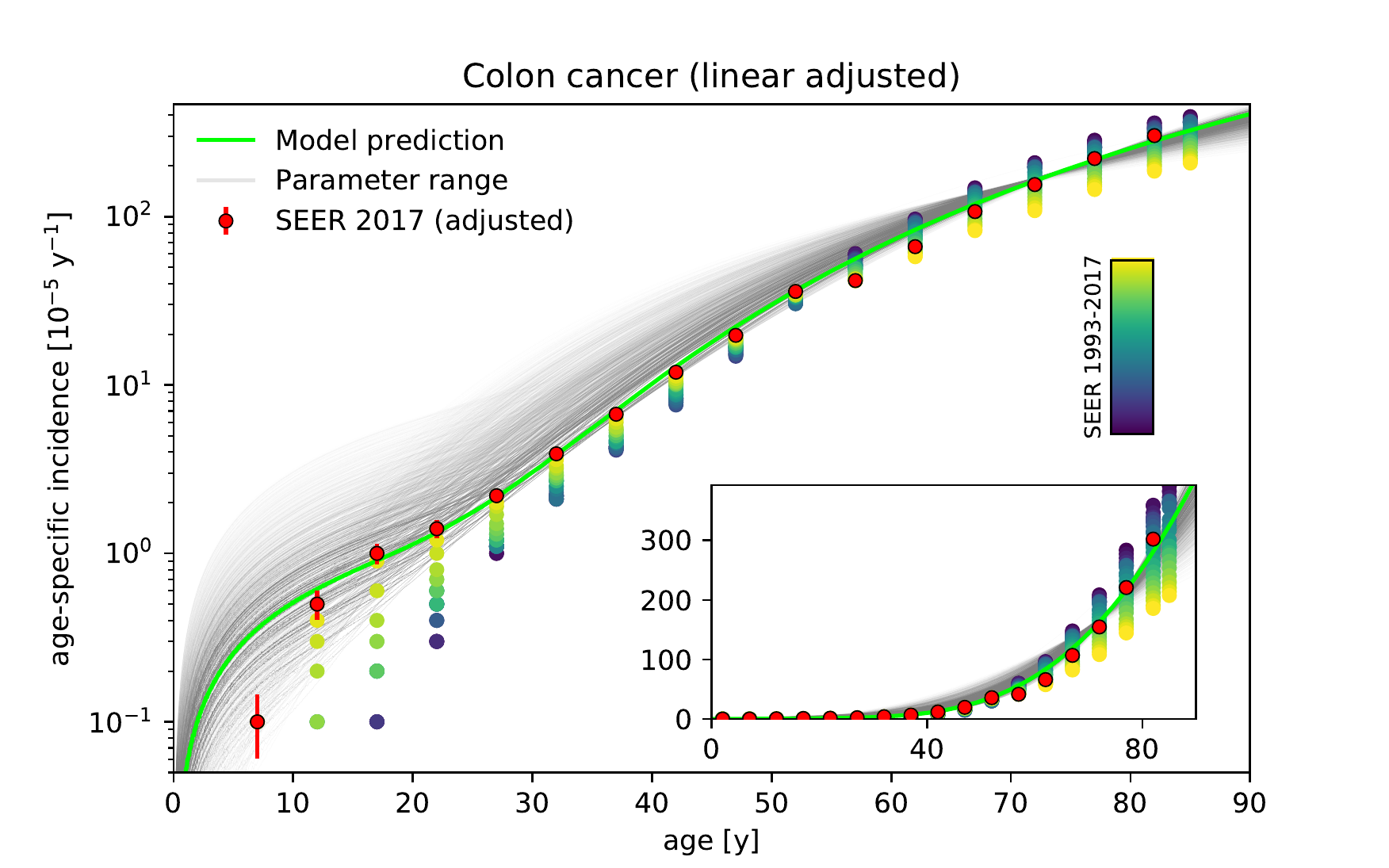}
\caption{\label{fig:colon-inc} \textbf{Model comparison to colon data
    adjusted for colorectal screening by incrementally increased
    rates.} Age-specific incidence rates of colon cancer predicted by
  the model and epidemiological data (SEER database~\cite{SEER2017}
  adjusted by incrementally increased rates $\geq 55$ years, see text
  for details, displayed curves and points analogous to
  \figref{incidence}). As illustration of secular trends, the
  epidemiological rates from the archive of the annual SEER
  database~\cite{SEERall} are displayed by points colored according to
  the release year of the corresponding report from 1993 (dark blue)
  to 2017 (yellow). Exemplary parameter set with $N = 9$,
  $\C = 1.5\cdot10^7$, $\mutb=4.4\cdot10^{-6}$,
  $\mutm = 1.75\cdot10^{-6}$, and $\pprog = 9.4\%$ is highlighted in
  green. The effective replacement rate $\rr$ covers a range
  $0.01-0.07 \text{ y}^{-1}$ per stem cell with an average
  $\rr = 0.025 \pm 0.01\text{ y}^{-1}$ ($\rr = 0.022 \text{ y}^{-1}$
  per stem cell for the green curve).}
\end{figure}

\begin{figure}[ht!]
  \centering
\hspace*{-1cm}\includegraphics{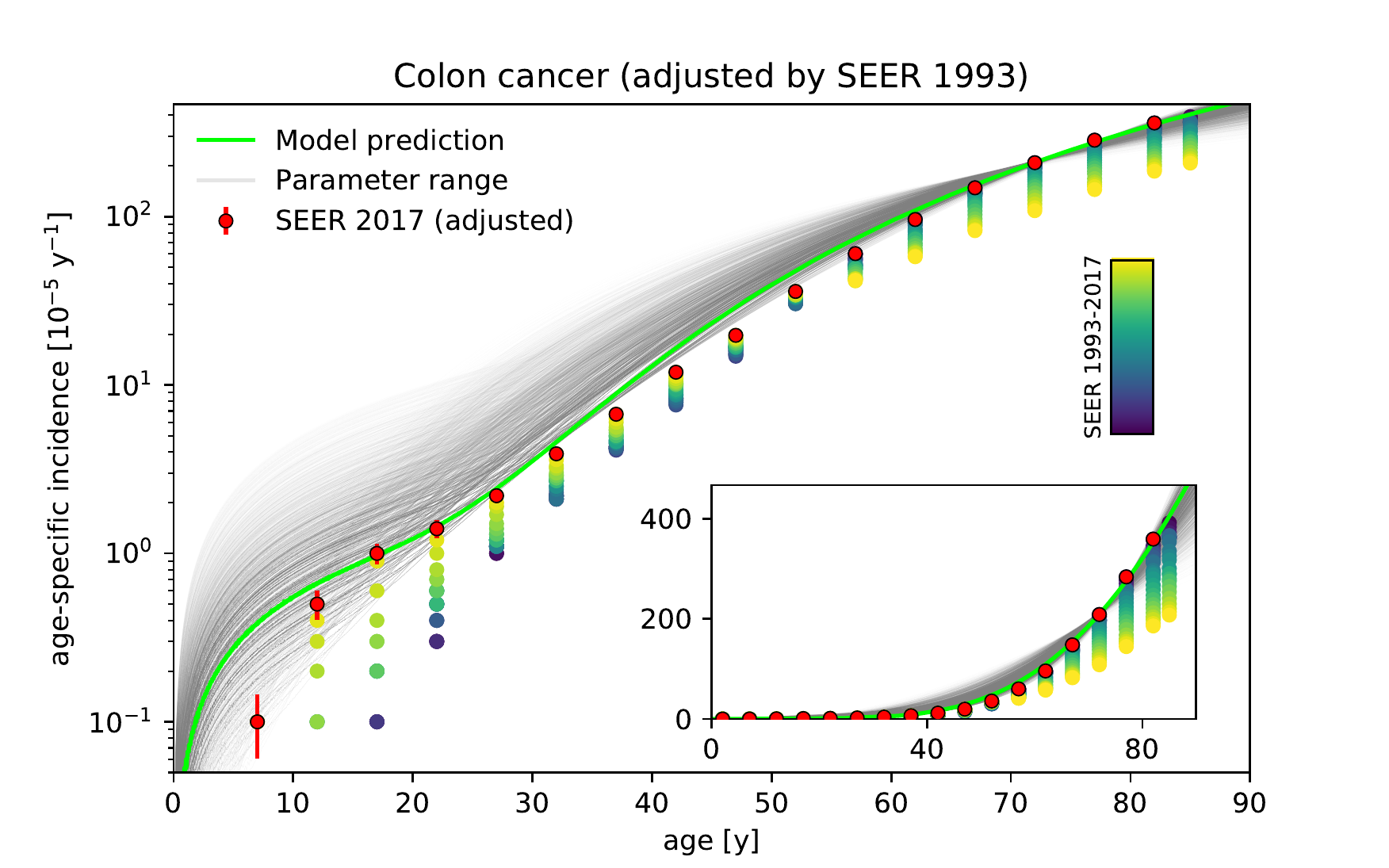}
\caption{\label{fig:colon-max} \textbf{Model comparison to colon data
    adjusted for colorectal screening by SEER 1993 database.}
  Age-specific incidence rates of colon cancer predicted by the model
  and epidemiological data (SEER database~\cite{SEER2017} adjusted by
  rates from SEER 1993~\cite{SEERall} for $\geq 55$ years, displayed
  curves and points analogous to \figref{incidence}). As illustration
  of secular trends, the epidemiological rates from the archive of the
  annual SEER database~\cite{SEERall} are displayed by points colored
  according to the release year of the corresponding report from 1993
  (dark blue) to 2017 (yellow). Exemplary parameter set with $N = 10$,
  $\C = 2\cdot10^7$, $\mutb=1.75\cdot10^{-6}$,
  $\mutm = 1.75\cdot10^{-6}$, and $\pprog = 9.4\%$ is highlighted in
  green. The effective replacement rate $\rr$ covers a range
  $0.01-0.09 \text{ y}^{-1}$ per stem cell with an average
  $\rr = 0.03 \pm 0.01\text{ y}^{-1}$ ($\rr = 0.03 \text{ y}^{-1}$ per
  stem cell for the green curve).}
\end{figure}

\begin{figure}[ht!]
  \centering
\hspace*{-1cm}\includegraphics{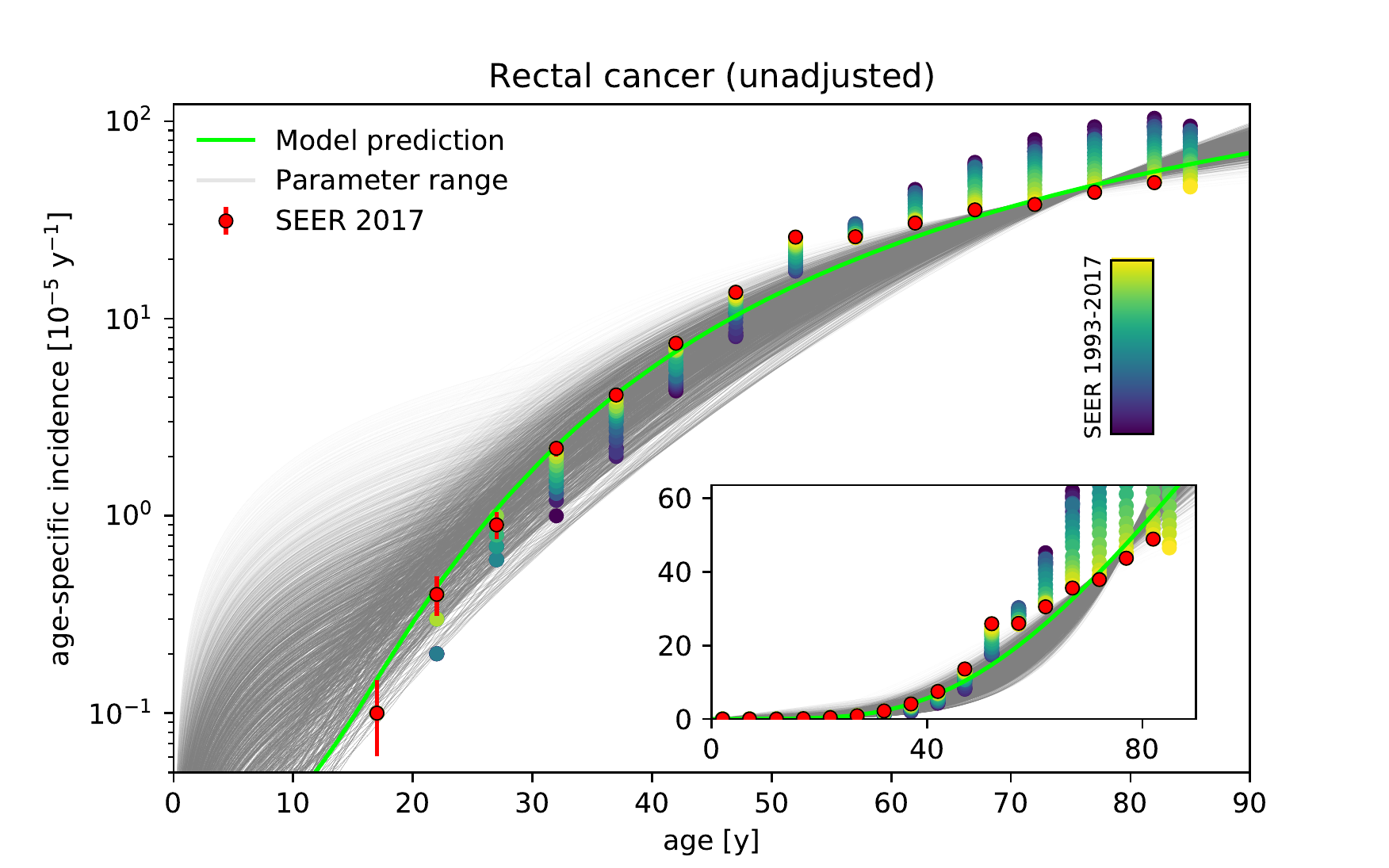}
\caption{\label{fig:rectum-no-adjust} \textbf{Model comparison to
    rectum data without adjustments for colorectal screening.}
  Age-specific incidence rates of rectal cancer predicted by the model
  and epidemiological data (SEER database~\cite{SEER2017} unadjusted,
  displayed curves and points analogous to \figref{incidence}). As
  illustration of secular trends, the epidemiological rates from the
  archive of the annual SEER database~\cite{SEERall} are displayed by
  points colored according to the release year of the corresponding
  report from 1993 (dark blue) to 2017 (yellow). Exemplary parameter
  set with $N = 8$, $\C = 10^6$, $\mutb=1.75\cdot10^{-6}$,
  $\mutm = 1.75\cdot10^{-6}$, and $\pprog = 5\%$ is highlighted in
  green. The effective replacement rate $\rr$ covers a range
  $0.01-0.16 \text{ y}^{-1}$ per stem cell with an average
  $\rr = 0.04 \pm 0.02\text{ y}^{-1}$ ($\rr = 0.036 \text{ y}^{-1}$
  per stem cell for the green curve).}
\end{figure}

\begin{figure}[ht!]
  \centering
\hspace*{-1cm}\includegraphics{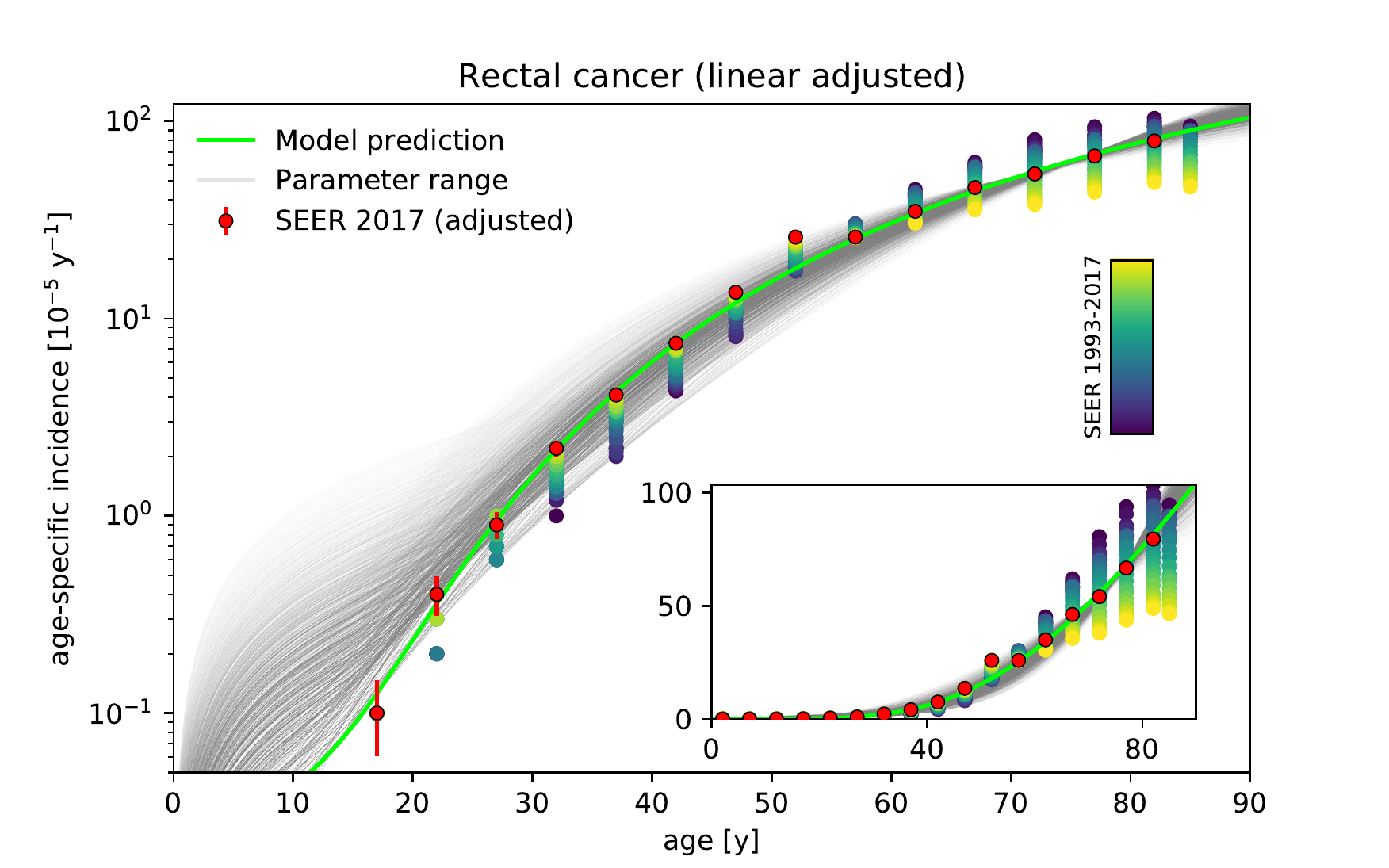}
\caption{\label{fig:rectum-inc} \textbf{Model comparison to rectum
    data adjusted for colorectal screening by incrementally increased
    rates.} Age-specific incidence rates of rectal cancer predicted by
  the model and epidemiological data (SEER database~\cite{SEER2017}
  adjusted by incrementally increased rates $\geq 55$ years, see text
  for details, displayed curves and points analogous to
  \figref{incidence}). As illustration of secular trends, the
  epidemiological rates from the archive of the annual SEER
  database~\cite{SEERall} are displayed by points colored according to
  the release year of the corresponding report from 1993 (dark blue)
  to 2017 (yellow). Exemplary parameter set with $N = 9$, $\C = 10^6$,
  $\mutb=1.75\cdot10^{-6}$, $\mutm = 1.75\cdot10^{-6}$, and
  $\pprog = 9.4\%$ is highlighted in green. The effective replacement
  rate $\rr$ covers a range $0.01-0.12 \text{ y}^{-1}$ per stem cell
  with an average $\rr = 0.04 \pm 0.02\text{ y}^{-1}$
  ($\rr = 0.038 \text{ y}^{-1}$ per stem cell for the green curve).}
\end{figure}

\begin{figure}[ht!]
  \centering
\hspace*{-1cm}\includegraphics{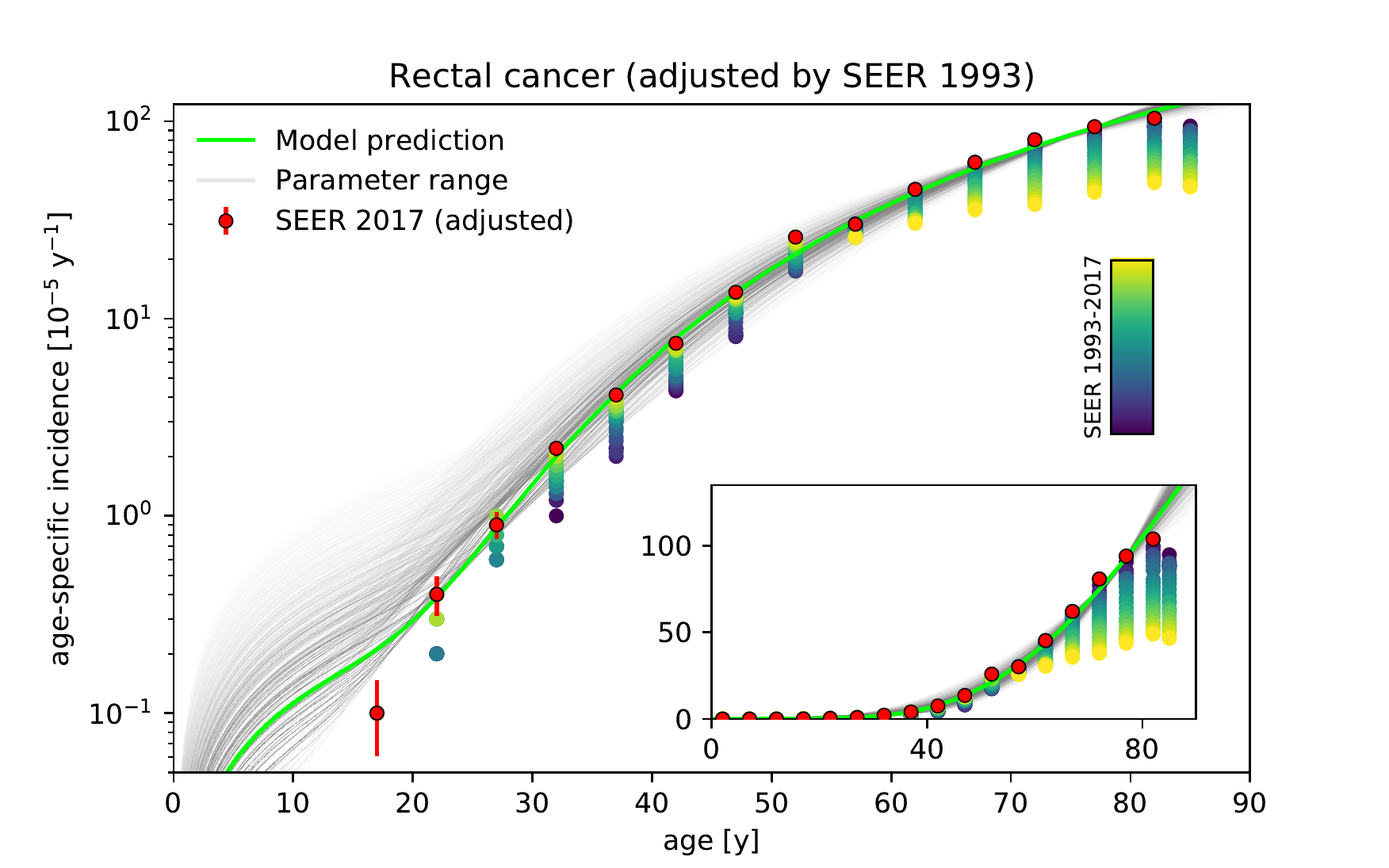}
\caption{\label{fig:rectum-max}\textbf{Model comparison to rectum data
    adjusted for colorectal screening by SEER 1993 database.}
  Age-specific incidence rates of rectal cancer predicted by the model
  and epidemiological data (SEER database~\cite{SEER2017} adjusted by
  rates from SEER 1993~\cite{SEERall} for $\geq 55$ years, displayed
  curves and points analogous to \figref{incidence}). As illustration
  of secular trends, the epidemiological rates from the archive of the
  annual SEER database~\cite{SEERall} are displayed by points colored
  according to the release year of the corresponding report from 1993
  (dark blue) to 2017 (yellow). Exemplary parameter set with $N = 12$,
  $\C = 10^6$, $\mutb=1.75\cdot10^{-6}$, $\mutm = 1.75\cdot10^{-6}$,
  and $\pprog = 9.4\%$ is highlighted in green. The effective
  replacement rate $\rr$ covers a range $0.01-0.12 \text{ y}^{-1}$ per
  stem cell with an average $\rr = 0.043 \pm 0.02\text{ y}^{-1}$
  ($\rr = 0.055 \text{ y}^{-1}$ per stem cell for the green curve).}
\end{figure}

While the exact quantification of the mentioned external effects is
not possible, we consider several reasonable adjustments of the
incidence rates from the SEER 2017 dataset to account for the effect
of colorectal cancer screening. First of all we present the original
incidences for colon and rectum from the SEER 2017 database in
\figref{colon-no-adjust} and \figref{rectum-no-adjust} along with the
incidences from the previous years $1993-2017$ for illustration of the
age-dependent secular trends. The first adjustment is based on
estimates that the annual incidence of colorectal cancer at ages $>50$
between 2000 and 2015 is reduced on average by $\sim 25\%$ due to
screening~\cite{Levetal2018}. Thus, the colon and rectal incidences
displayed \figref{incidence} and \figref{rectal-cancer} correspond to
the SEER database of 2017, for which the rates of all age groups above
$55$ years have been increased by a factor $4/3$. The second
adjustment is based on the estimate that the impact of screening
colonoscopy is modest in the age group $55-64$ and increases for older
age groups~\cite{BreAltStoHof2015}. From the predictions of
Ref.~\cite{BreAltStoHof2015} we obtain roughly an $8\%$ reduction of
annual incidences for the age group $55-64$, $28\%$ for age group
$65-74$, and $37\%$ for the age group $75-84$. We assign these
percentages to the ages $60$, $70$, and $80$, respectively, and
linearly interpolate the annual reductions between these ages, which
we use to again adjust the results of the SEER database of 2017 in
\figref{colon-inc} and \figref{rectum-inc}. Finally, as a limit case
we set the incidences above $55$ year to the ones reported in the SEER
1993 database, see \figref{colon-max} and \figref{rectum-max}.

For the colon, we find for each adjustment (and without adjustment) a
similar correspondence between the epidemiological incidence rates and
our model prediction. Mostly, the fitted effective replacement rate
for a particular parameter set is modulated by the choice of the
adjustment. For the rectum, the model displays a better correspondence
with the incidence rates strongly adjusted for screening effects, see
\figref{rectum-inc} and \figref{rectum-max}. Note that the risk
reduction is known to vary by subsite of the colon and
rectum~\cite{BreAltStoHof2015} while our estimates of this risk
reduction are based on colorectal
cancer~\cite{KahImpJulRex2009,BreAltStoHof2015,Levetal2018}, whose
incidences are dominated by incidences of colon cancer.

Note that at ages younger than $50$ colorectal screening occurs rarely
and instead the rates may be subject to two opposing effects: On the
one hand, the apparent rise in incidences below $50$ years may be
attributed to life-style
changes~\cite{SieDeSJem2014,SieFedAndMilMaRosJem2017} an effect not
incorporated in our biological model. On the other hand, recent
results support the presence of a large undetected preclinical case
burden $<50$ years, which is not reflected in the rates of colorectal
cancer observed in the SEER registries~\cite{AbuZhoAhnQinXiaKar2020}.
The latter is consistent with the fact, that people younger than $55$
are more likely to be diagnosed with late-stage disease, largely due
to sometimes for years delayed follow-up of
symptoms~\cite{SieFedAndMilMaRosJem2017}. This implies that cancer
incidences are either missed at younger ages or wrongfully assigned to
older age groups, whose rates they barely impact as the age-specific
rates increase significantly with age. Since the net impact of these
two opposing effects is unkown, we do not adjust incidence rates below
$55$ years, similar to Ref.~\cite{LueCurJeoHaz2013}.

\begin{figure}[ht!]
  \centering
\hspace*{-1cm}\includegraphics{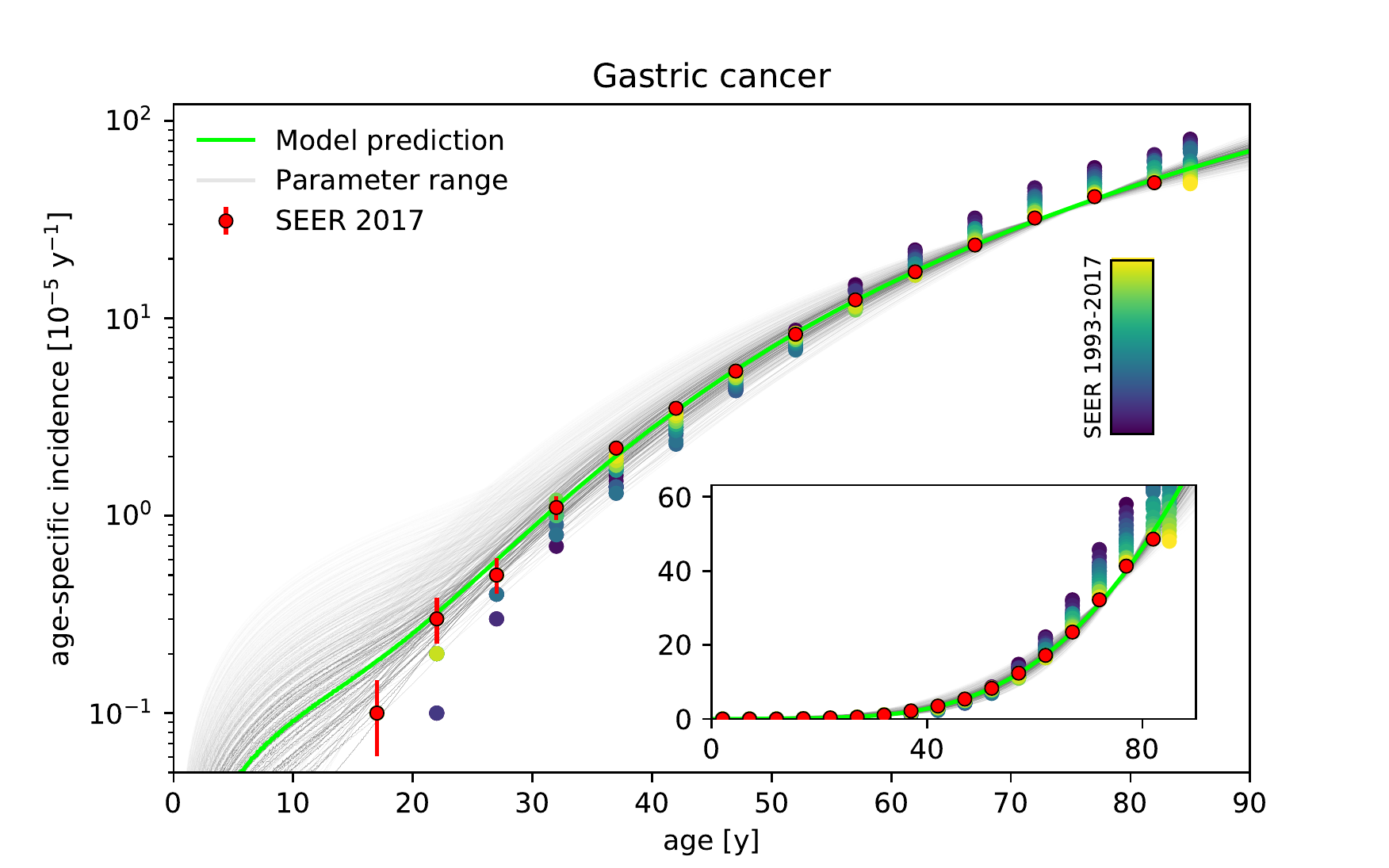}
\caption{\label{fig:stomach-no-adjust} \textbf{Relation to trends over
    calendar year for gastric cancer.} Age-specific incidence rates of
  gastric cancer predicted from the model and the epidemiological data
  as in \figref{gastric-cancer}. As illustration of secular trends,
  the epidemiological rates from the archive of the annual SEER
  database~\cite{SEERall} are displayed by points colored according to
  the release year of the corresponding report from 1993 (dark blue)
  to 2017 (yellow). While qualitatively the same trends at younger and
  older age as for colon and rectal cancer are visible, these trends
  are significantly smaller for gastric cancer.}
\end{figure}

Note that the secular trends for gastric cancer are less pronounced
compared to colorectal cancer, see \figref{stomach-no-adjust}, and
thus are used unadjusted from the SEER 2017 database.

\clearpage

\begin{thebibliography}{10}

\bibitem{VogKin1993}
Vogelstein B, Kinzler KW.
\newblock The multistep nature of cancer.
\newblock Trends in Genetics. 1993;9(4):138--141.
\newblock doi:{10.1016/0168-9525(93)90209-Z}.

\bibitem{LitVinLi2008}
Little MP, Vineis P, Li G.
\newblock A stochastic carcinogenesis model incorporating multiple types of
  genomic instability fitted to colon cancer data.
\newblock J~Theor~Biol. 2008;254(2):229--238.
\newblock doi:{10.1016/j.jtbi.2008.05.027}.

\bibitem{FeaHamVog1987}
Fearon ER, Hamilton, Vogelstein B.
\newblock Clonal analysis of human colorectal tumors.
\newblock Science. 1987;238(4824):193--197.
\newblock doi:{10.1126/science.2889267}.

\bibitem{ReyMorClaWei2001}
Reya T, Morrison SJ, Clarke MF, Weissman IL.
\newblock Stem cells, cancer, and cancer stem cells.
\newblock Nature. 2001;414(6859):105--111.
\newblock doi:{10.1038/35102167}.

\bibitem{WilWerBarGraSot2016}
Williams MJ, Werner B, Barnes CP, Graham TA, Sottoriva A.
\newblock Identification of neutral tumor evolution across cancer types.
\newblock Nature Genetics. 2016;48(3):238--244.
\newblock doi:{10.1038/ng.3489}.

\bibitem{WuWanLinLu2016}
Wu CI, Wang HY, Ling S, Lu X.
\newblock The Ecology and Evolution of Cancer: The Ultra-Microevolutionary
  Process.
\newblock Annu~Rev~Genet. 2016;50(1):347--369.
\newblock doi:{10.1146/annurev-genet-112414-054842}.

\bibitem{AmoBac2014}
Amoyel M, Bach EA.
\newblock Cell competition: how to eliminate your neighbours.
\newblock Development. 2014;141(5):988--1000.
\newblock doi:{10.1242/dev.079129}.

\bibitem{LodBerZipMatBalDar2000}
Lodish H, Berk A, Zipursky SL, Matsudaira P, Baltimore D, Darnell J.
\newblock Tumor Cells and the Onset of Cancer.
\newblock Molecular Cell Biology 4th edition. 2000;.

\bibitem{KuaNagEik2016}
Kuang Y, Nagy JD, Eikenberry SE.
\newblock Introduction to Mathematical Oncology.
\newblock {CRC} Press; 2016.

\bibitem{FoxWan2007}
Fox JG, Wang TC.
\newblock Inflammation, atrophy, and gastric cancer.
\newblock J~Clin~Invest. 2007;117(1):60--69.
\newblock doi:{10.1172/JCI30111}.

\bibitem{ZeuTodStaDeM2014}
Zeuner A, Todaro M, Stassi G, De~Maria R.
\newblock Colorectal Cancer Stem Cells: From the Crypt to the Clinic.
\newblock Cell~Stem~Cell. 2014;15(6):692--705.
\newblock doi:{10.1016/j.stem.2014.11.012}.

\bibitem{Sotetal2015}
Sottoriva A, Kang H, Ma Z, Graham TA, Salomon MP, Zhao J, et~al.
\newblock A Big Bang model of human colorectal tumor growth.
\newblock Nature Genetics. 2015;47(3):209--216.
\newblock doi:{10.1038/ng.3214}.

\bibitem{ArmDol1954}
Armitage P, Doll R.
\newblock The Age Distribution of Cancer and a Multi-stage Theory of
  Carcinogenesis.
\newblock Br~J~Cancer. 1954;8(1):1--12.
\newblock doi:{10.1038/bjc.1954.1}.

\bibitem{ArmDol2004}
Armitage P, Doll R.
\newblock The age distribution of cancer and a multi-stage theory of
  carcinogenesis.
\newblock Br~J~Cancer. 2004;91(12):1983--1989.
\newblock doi:{10.1038/sj.bjc.6602297}.

\bibitem{MooMezTur2009}
Moolgavkar SH, Meza R, Turim J.
\newblock Pleural and peritoneal mesotheliomas in SEER: age effects and
  temporal trends, 1973-2005.
\newblock Cancer Causes Control. 2009;20(6):935--944.
\newblock doi:{10.1007/s10552-009-9328-9}.

\bibitem{MezJeoMooLue2008}
Meza R, Jeon J, Moolgavkar SH, Luebeck EG.
\newblock Age-specific incidence of cancer: Phases, transitions, and biological
  implications.
\newblock Proc~Natl~Acad~Sci~USA. 2008;105(42):16284--16289.
\newblock doi:{10.1073/pnas.0801151105}.

\bibitem{MezJeoRenLue2010}
Meza R, Jeon J, Renehan AG, Luebeck EG.
\newblock Colorectal Cancer Incidence Trends in the United States and United
  Kingdom: Evidence of Right- to Left-Sided Biological Gradients with
  Implications for Screening.
\newblock Cancer~Res. 2010;70(13):5419--5429.
\newblock doi:{10.1158/0008-5472.CAN-09-4417}.

\bibitem{LueCurJeoHaz2013}
Luebeck EG, Curtius K, Jeon J, Hazelton WD.
\newblock Impact of Tumor Progression on Cancer Incidence Curves.
\newblock Cancer~Res. 2013;73(3):1086--1096.
\newblock doi:{10.1158/0008-5472.CAN-12-2198}.

\bibitem{MezCha2015}
Meza R, Chang JT.
\newblock Multistage carcinogenesis and the incidence of thyroid cancer in the
  US by sex, race, stage and histology.
\newblock BMC~Public Health. 2015;15(1):789.
\newblock doi:{10.1186/s12889-015-2108-4}.

\bibitem{BroEisMez2016}
Brouwer AF, Eisenberg MC, Meza R.
\newblock Age Effects and Temporal Trends in HPV-Related and HPV-Unrelated Oral
  Cancer in the United States: A Multistage Carcinogenesis Modeling Analysis.
\newblock PLoS ONE. 2016;11(3):e0151098.
\newblock doi:{10.1371/journal.pone.0151098}.

\bibitem{BroMezEis2017}
Brouwer AF, Meza R, Eisenberg MC.
\newblock Parameter estimation for multistage clonal expansion models from
  cancer incidence data: A practical identifiability analysis.
\newblock PLoS~Comput~Biol. 2017;13(3):e1005431.
\newblock doi:{10.1371/journal.pcbi.1005431}.

\bibitem{CalTavShi2004}
Calabrese P, Tavar\'{e} S, Shibata D.
\newblock Pretumor Progression: Clonal Evolution of Human Stem Cell
  Populations.
\newblock Am~J~Pathol. 2004;164(4):1337--1346.
\newblock doi:{10.1016/S0002-9440(10)63220-8}.

\bibitem{KimCalTavShi2004}
Kim KM, Calabrese P, Tavar\'{e} S, Shibata D.
\newblock Enhanced Stem Cell Survival in Familial Adenomatous Polyposis.
\newblock Am~J~Pathol. 2004;164(4):1369--1377.
\newblock doi:{10.1016/S0002-9440(10)63223-3}.

\bibitem{LanKuiMisBee2020}
Lang BM, Kuipers J, Misselwitz B, Beerenwinkel N.
\newblock Predicting colorectal cancer risk from adenoma detection via a
  two-type branching process model.
\newblock PLoS~Comput~Biol. 2020;16(2):e1007552.
\newblock doi:{10.1371/journal.pcbi.1007552}.

\bibitem{GroBraHolHauKunTreAalMog2011}
Grotmol T, Bray F, Holte H, Haugen M, Kunz L, Tretli S, et~al.
\newblock Frailty Modeling of the Bimodal Age-Incidence of Hodgkin Lymphoma in
  the Nordic Countries.
\newblock Cancer~Epidem~Biomar. 2011;20(7):1350--1357.
\newblock doi:{10.1158/1055-9965.EPI-10-1014}.

\bibitem{MdzGleKinShe2009}
Mdzinarishvili T, Gleason MX, Kinarsky L, Sherman S.
\newblock A Generalized Beta Model for the Age Distribution of Cancers:
  Application to Pancreatic and Kidney Cancer.
\newblock Cancer~Inform. 2009;7:CIN.S3050.
\newblock doi:{10.4137/CIN.S3050}.

\bibitem{MdzShe2010}
Mdzinarishvili T, Sherman S.
\newblock Weibull-like Model of Cancer Development in Aging.
\newblock Cancer~Inform. 2010;9:CIN.S5460.
\newblock doi:{10.4137/CIN.S5460}.

\bibitem{Bro2011}
Brody JP.
\newblock Age-Specific Incidence Data Indicate Four Mutations Are Required for
  Human Testicular Cancers.
\newblock PLoS ONE. 2011;6(10):e25978.
\newblock doi:{10.1371/journal.pone.0025978}.

\bibitem{SotBro2012}
Soto-Ortiz L, Brody JP.
\newblock A theory of the cancer age-specific incidence data based on extreme
  value distributions.
\newblock AIP~Adv. 2012;2(1):011205.
\newblock doi:{10.1063/1.3699050}.

\bibitem{RhyOhKimKimRhyHon2021}
Rhyu MG, Oh JH, Kim TH, Kim JS, Rhyu YA, Hong SJ.
\newblock Periodic Fluctuations in the Incidence of Gastrointestinal Cancer.
\newblock Front~Oncol. 2021;11.
\newblock doi:{10.3389/fonc.2021.558040}.

\bibitem{NobBurLeLemVioKatBee2022}
Noble R, Burri D, Le~Sueur C, Lemant J, Viossat Y, Kather JN, et~al.
\newblock Spatial structure governs the mode of tumour evolution.
\newblock Nat~Ecol~Evol. 2022;6(2):207--217.
\newblock doi:{10.1038/s41559-021-01615-9}.

\bibitem{PatCleBoz2020}
Paterson C, Clevers H, Bozic I.
\newblock Mathematical model of colorectal cancer initiation.
\newblock Proc~Natl~Acad~Sci~USA. 2020;117(34):20681--20688.
\newblock doi:{10.1073/pnas.2003771117}.

\bibitem{SEER2017}
Surveillance, Epidemiology, and End Results (SEER) Program
  (www.seer.cancer.gov) Research Data (1975-2017), National Cancer Institute,
  DCCPS, Surveillance Research Program, released April 2020, based on the
  November 2019 submission.
  \url{https://seer.cancer.gov/archive/csr/1975_2017/download_csr_datafile.php/sect_06_table.11.csv};
  \url{https://seer.cancer.gov/archive/csr/1975_2017/download_csr_datafile.php/sect_24_table.07.csv};.

\bibitem{BudDeuKliVos2019}
Buder T, Deutsch A, Klink B, Voss-B\"ohme A.
\newblock Patterns of Tumor Progression Predict Small and Tissue-Specific
  Tumor-Originating Niches.
\newblock Front~Oncol. 2019;8.
\newblock doi:{10.3389/fonc.2018.00668}.

\bibitem{DriRid1994}
Driman DK, Riddell RH.
\newblock Flat adenomas and flat carcinomas: Do you see what I see?
\newblock Gastrointest~Endosc. 1994;40(1):106--109.
\newblock doi:{10.1016/S0016-5107(94)70028-1}.

\bibitem{SEERpop}
United States Standard Population 2000 used by SEER based on the Census
  P25-1130
  \url{https://seer.cancer.gov/csr/1975_2018/browse_csr.php?sectionSEL=6&pageSEL=sect_a_table.08}
  (Day, Jennifer Cheeseman, Population Projections of the United States by Age,
  Sex, Race, and Hispanic Origin: 1995 to 2050, U.S. Bureau of the Census,
  Current Population Reports, P25-1130, U.S. Government Printing Office,
  Washington, DC, 1996);.

\bibitem{SieDeSJem2014}
Siegel R, DeSantis C, Jemal A.
\newblock Colorectal cancer statistics, 2014.
\newblock CA:~Cancer~J~Clin. 2014;64(2):104--117.
\newblock doi:{10.3322/caac.21220}.

\bibitem{Baietal2015}
Bailey CE, Hu CY, You YN, Bednarski BK, Rodriguez-Bigas MA, Skibber JM, et~al.
\newblock Increasing Disparities in the Age-Related Incidences of Colon and
  Rectal Cancers in the United States, 1975-2010.
\newblock JAMA~Surg. 2015;150(1):17--22.
\newblock doi:{10.1001/jamasurg.2014.1756}.

\bibitem{KahImpJulRex2009}
Kahi CJ, Imperiale TF, Juliar BE, Rex DK.
\newblock Effect of Screening Colonoscopy on Colorectal Cancer Incidence and
  Mortality.
\newblock Clin~Gastroenterol~Hepatol. 2009;7(7):770--775.
\newblock doi:{10.1016/j.cgh.2008.12.030}.

\bibitem{Levetal2018}
Levin TR, Corley DA, Jensen CD, Schottinger JE, Quinn VP, Zauber AG, et~al.
\newblock Effects of Organized Colorectal Cancer Screening on Cancer Incidence
  and Mortality in a Large Community-Based Population.
\newblock Gastroenterology. 2018;155(5):1383--1391.e5.
\newblock doi:{10.1053/j.gastro.2018.07.017}.

\bibitem{CarZhuGuoHeiHofBre2021}
Cardoso R, Zhu A, Guo F, Heisser T, Hoffmeister M, Brenner H.
\newblock Incidence and Mortality of Proximal and Distal Colorectal Cancer in
  Germany.
\newblock Dtsch~Arztebl~Int. 2021;118(16):281--287.
\newblock doi:{10.3238/arztebl.m2021.0111}.

\bibitem{BreAltStoHof2015}
Brenner H, Altenhofen L, Stock C, Hoffmeister M.
\newblock Expected long-term impact of the German screening colonoscopy
  programme on colorectal cancer prevention: Analyses based on 4,407,971
  screening colonoscopies.
\newblock Eur~J~Cancer. 2015;51(10):1346--1353.
\newblock doi:{10.1016/j.ejca.2015.03.020}.

\bibitem{SieFedAndMilMaRosJem2017}
Siegel RL, Fedewa SA, Anderson WF, Miller KD, Ma J, Rosenberg PS, et~al.
\newblock Colorectal Cancer Incidence Patterns in the United States, 1974-2013.
\newblock J~Natl~Cancer~Inst. 2017;109(8):djw322.
\newblock doi:{10.1093/jnci/djw322}.

\bibitem{AbuZhoAhnQinXiaKar2020}
Abualkhair WH, Zhou M, Ahnen D, Yu Q, Wu XC, Karlitz JJ.
\newblock Trends in Incidence of Early-Onset Colorectal Cancer in the United
  States Among Those Approaching Screening Age.
\newblock JAMA~Netw~Open. 2020;3(1):e1920407.
\newblock doi:{10.1001/jamanetworkopen.2019.20407}.

\bibitem{LueMoo2002}
Luebeck EG, Moolgavkar SH.
\newblock Multistage carcinogenesis and the incidence of colorectal cancer.
\newblock Proc~Natl~Acad~Sci~USA. 2002;99(23):15095--15100.
\newblock doi:{10.1073/pnas.222118199}.

\bibitem{HumWri2008}
Humphries A, Wright NA.
\newblock Colonic crypt organization and tumorigenesis.
\newblock Nat~Rev~Cancer. 2008;8(6):415--424.
\newblock doi:{10.1038/nrc2392}.

\bibitem{BudDeuKliVos2015}
Buder T, Deutsch A, Klink B, Voss-B\"ohme A.
\newblock Model-Based Evaluation of Spontaneous Tumor Regression in Pilocytic
  Astrocytoma.
\newblock PLoS~Comput~Biol. 2015;11(12):e1004662.
\newblock doi:{10.1371/journal.pcbi.1004662}.

\bibitem{VogKin2002}
Vogelstein B, Kinzler KW.
\newblock The Genetic Basis of Human Cancer.
\newblock Subsequent edition ed. Mcgraw-Hill Professional; 2002.

\bibitem{DurMos2015}
Durrett R, Moseley S.
\newblock Spatial Moran models I. Stochastic tunneling in the neutral case.
\newblock Ann~Appl~Prob. 2015;25(1):104--115.
\newblock doi:{10.1214/13-AAP989}.

\bibitem{Mesetal2018}
Mesa KR, Kawaguchi K, Cockburn K, Gonzalez D, Boucher J, Xin T, et~al.
\newblock Homeostatic Epidermal Stem Cell Self-Renewal Is Driven by Local
  Differentiation.
\newblock Cell~Stem~Cell. 2018;23(5):677--686.e4.
\newblock doi:{10.1016/j.stem.2018.09.005}.

\bibitem{PagRadRepHas2015}
Paggi S, Radaelli F, Repici A, Hassan C.
\newblock Advances in the removal of diminutive colorectal polyps.
\newblock Expert~Rev~Gastroenterol~Hepatol. 2015;9(2):237--244.
\newblock doi:{10.1586/17474124.2014.950955}.

\bibitem{Nicetal2018}
Nicholson AM, Olpe C, Hoyle A, Thorsen AS, Rus T, Colomb\'{e} M, et~al.
\newblock Fixation and Spread of Somatic Mutations in Adult Human Colonic
  Epithelium.
\newblock Cell~Stem~Cell. 2018;22(6):909--918.e8.
\newblock doi:{10.1016/j.stem.2018.04.020}.

\bibitem{Baketal2014}
Baker AM, Cereser B, Melton S, Fletcher AG, Rodriguez-Justo M, Tadrous PJ,
  et~al.
\newblock Quantification of Crypt and Stem Cell Evolution in the Normal and
  Neoplastic Human Colon.
\newblock Cell~Rep. 2014;8(4):940--947.
\newblock doi:{10.1016/j.celrep.2014.07.019}.

\bibitem{Gabetal2022}
Gabbutt C, Schenck RO, Weisenberger DJ, Kimberley C, Berner A, Househam J,
  et~al.
\newblock Fluctuating methylation clocks for cell lineage tracing at high
  temporal resolution in human tissues.
\newblock Nat~Biotechnol. 2022; p. 1--11.
\newblock doi:{10.1038/s41587-021-01109-w}.

\bibitem{TomVog2015}
Tomasetti C, Vogelstein B.
\newblock Variation in cancer risk among tissues can be explained by the number
  of stem cell divisions.
\newblock Science. 2015;347(6217):78--81.
\newblock doi:{10.1126/science.1260825}.

\bibitem{HouDesDesBerVel2002}
Hounnou G, Destrieux C, Desm\'{e} J, Bertrand P, Velut S.
\newblock Anatomical study of the length of the human intestine.
\newblock Surg~Radiol~Anat. 2002;24(5):290--294.
\newblock doi:{10.1007/s00276-002-0057-y}.

\bibitem{GraVar2008}
Grahn SW, Varma MG.
\newblock Factors that increase risk of colon polyps.
\newblock Clin~Colon~Rectal~Surg. 2008;21(4):247--255.
\newblock doi:{10.1055/s-0028-1089939}.

\bibitem{Steetal2013}
Steele SR, Johnson EK, Champagne B, Davis B, Lee S, Rivadeneira D, et~al.
\newblock Endoscopy and polyps-diagnostic and therapeutic advances in
  management.
\newblock World Journal of Gastroenterology. 2013;19(27):4277--4288.
\newblock doi:{10.3748/wjg.v19.i27.4277}.

\bibitem{SieMarWooTavShi2009}
Siegmund KD, Marjoram P, Woo YJ, Tavar\'{e} S, Shibata D.
\newblock Inferring clonal expansion and cancer stem cell dynamics from DNA
  methylation patterns in colorectal cancers.
\newblock Proc~Natl~Acad~Sci~USA. 2009;106(12):4828--4833.
\newblock doi:{10.1073/pnas.0810276106}.

\bibitem{BedFaiBouPiaCauHil1992}
Bedenne L, Faivre J, Boutron MC, Piard F, Cauvin JM, Hillon P.
\newblock Adenoma-carcinoma sequence or “de novo” Carcinogenesis?. A study
  of adenomatous remnants in a population-based series of large bowel cancers.
\newblock Cancer. 1992;69(4):883--888.
\newblock
  doi:{https://doi.org/10.1002/1097-0142(19920215)69:4<883::AID-CNCR2820690408>3.0.CO;2-B}.

\bibitem{MeiMorMieBaySto2001}
Meining A, Morgner A, Miehlke S, Bayerd\"{o}rffer E, Stolte M.
\newblock Atrophy–metaplasia–dysplasia–carcinoma sequence in the stomach:
  a reality or merely an hypothesis?
\newblock Best~Pract~Res~Clin~Gastroenterol. 2001;15(6):983--998.
\newblock doi:{10.1053/bega.2001.0253}.

\bibitem{LopKleSimWin2010}
Lopez-Garcia C, Klein AM, Simons BD, Winton DJ.
\newblock Intestinal Stem Cell Replacement Follows a Pattern of Neutral Drift.
\newblock Science. 2010;330(6005):822--825.
\newblock doi:{10.1126/science.1196236}.

\bibitem{HayFoxWan2017}
Hayakawa Y, Fox JG, Wang TC.
\newblock The Origins of Gastric Cancer From Gastric Stem Cells: Lessons From
  Mouse Models.
\newblock Cell~Mol~Gastroenterol~Hepatol. 2017;3(3):331--338.
\newblock doi:{10.1016/j.jcmgh.2017.01.013}.

\bibitem{KurKurDmiVce2001}
Kurbel S, Kurbel B, Dmitrovi\'{c} B, V\v{c}ev A.
\newblock A Model of the Gastric Gland Ejection Cycle: Low Ejection Fractions
  Require Reduction of the Glandular Dead Space.
\newblock J~Theor~Biol. 2001;210(3):337--343.
\newblock doi:{10.1006/jtbi.2001.2313}.

\bibitem{OhgKle2013}
Ohgaki H, Kleihues P.
\newblock The Definition of Primary and Secondary Glioblastoma.
\newblock Clin~Cancer~Res. 2013;19(4):764--772.
\newblock doi:{10.1158/1078-0432.CCR-12-3002}.

\bibitem{VijElaKha2015}
Vijay A, Elaffandi A, Khalaf H.
\newblock Hepatocellular adenoma: An update.
\newblock World~J~Hepatol. 2015;7(25):2603--2609.
\newblock doi:{10.4254/wjh.v7.i25.2603}.

\bibitem{Kom2006}
Komarova NL.
\newblock Spatial Stochastic Models for Cancer Initiation and Progression.
\newblock Bull~Math~Biol. 2006;68(7):1573--1599.
\newblock doi:{10.1007/s11538-005-9046-8}.

\bibitem{NewSteAllIng2014}
Newville M, Stensitzki T, Allen DB, Ingargiola A. LMFIT: Non-Linear
  Least-Square Minimization and Curve-Fitting for Python; 2014.
\newblock Available from: \url{https://zenodo.org/record/11813}.

\bibitem{SEERall}
Surveillance, Epidemiology, and End Results (SEER) Program
  (www.seer.cancer.gov) Cancer Statistics Review Archive (1993-2017), National
  Cancer Institute, DCCPS, Surveillance Research Program
  \url{https://seer.cancer.gov/csr/previous.html};.

\end{thebibliography}
\end{document}